\documentclass[12pt, floatfix, superscriptaddress]{revtex4}

\usepackage{amsmath}
\usepackage{amssymb}
\usepackage{amsfonts}   
\usepackage{graphicx}   
\usepackage{verbatim}   
\usepackage{color}      
\usepackage{subfigure}
\usepackage{pgfplotstable}

\usepackage{tikz}
\usetikzlibrary{positioning}
\definecolor{myRed}{RGB}{229,25,50}
\definecolor{myBlue}{RGB}{25,178,255}
\definecolor{myGreen}{RGB}{50,255,0}

\newcommand{\unit}[1]{\ensuremath{\, \mathrm{#1}}}

\begin{document}

\title{Loopless non-trapping invasion percolation model for fracking}
\date{\today}

\author{J. Quinn Norris}
\email{jqnorris@ucdavis.edu}
\affiliation{Department of Physics, One Shields Ave., University of California, Davis, CA 95616, United States}
\author{Donald L. Turcotte}
\email{dlturcotte@ucdavis.edu}
\affiliation{Department of Geology, One Shields Ave., University of California, Davis, CA 95616, United States}
\author{John B. Rundle}
\email{rundle@physics.ucdavis.edu}
\affiliation{Department of Physics, One Shields Ave., University of California, Davis, CA 95616, United States}
\affiliation{Department of Geology, One Shields Ave., University of California, Davis, CA 95616, United States}
\affiliation{Santa Fe Institute, Santa Fe, NM 87501, United States}

\date{\today}
\begin{abstract}
Recent developments in hydraulic fracturing (fracking) have enabled the recovery of large quantities of natural gas and oil from old, low permeability shales. These developments include a change from low-volume, high-viscosity fluid injection to high-volume, low-viscosity injection. The injected fluid introduces distributed damage that provides fracture permeability for the extraction of the gas and oil. In order to model this process, we utilize a loopless non-trapping invasion percolation previously introduced to model optimal polymers in a strongly disordered medium, and for determining minimum energy spanning trees on a lattice. We performed numerical simulations on a 2D square lattice and find significant differences from other percolation models. Additionally, we find that the growing fracture network satisfies both Horton-Strahler and Tokunaga network statistics. As with other invasion percolation models, our model displays burst dynamics, in which the cluster extends rapidly into a connected region. We introduce an alternative definition of bursts to be a consecutive series of opened bonds whose strengths are all below a specified value. Using this definition of bursts, we find good agreement with a power-law frequency-area distribution. These results are generally consistent with the observed distribution of microseismicity observed during a high-volume frack.
\end{abstract}

\maketitle

\section{Introduction}
Fracking is the use of hydraulic fractures to enhance the permeability in petroleum reservoirs. The fluid is water with small quantities of additives. In a frack, water is injected from a well perforation that can be regarded as a point source. The objective is to generate fractures that result in distributed damage and a network of flow pathways to the well after the frack is completed. In this paper we utilize a loopless non-trapping invasion percolation model for the statistical migration of an invading fluid from a point source. We hypothesize that the migrating fluid reactivates preexisting healed fractures.

A standard procedure in oil (or gas) production is to inject a fluid (typically water) in injection wells in order to drive the oil (or gas) to production wells. In order to model this process, the invasion percolation model was introduced \cite{Wilkinson1983}. In two dimensions a square grid of sites represent the fluid filled pores. The sites are connected by bonds through which fluid can flow. In this model all sites are initially occupied by a defending fluid, for example oil. An invading fluid such as water is injected along one side of the region under consideration; typically, this region is a layer of width $L$. The bonds are assigned random strengths $s$. A random number is drawn from a uniform distribution in the range $[0,1]$. At each time step, the invading fluid flows through the weakest pathway (smallest $s$) displacing the defending fluid. As time proceeds the cluster of occupied invading sites grows from the injection side of the layer, the maximum selected strength $s$ also increases with time. When the cluster crosses the region in consideration, it has been found to follow many power-law scalings \cite{Furuberg1988}. Additionally, the highest strength of the invaded bonds is approximately equal to the critical value $p_{c}$ of the static model for bond percolation.

The original version of the invasion percolation model assumed that the defending fluid (oil) was incompressible. Thus when a region became disconnected from an edge of the lattice, that region could no longer be invaded and the defending fluid was trapped. One of the earliest and simplest variants was to remove trapping rule. Models without the trapping rule are called non-trapping or compressible invasion percolation. It was found that the non-trapping version of invasion percolation is equivalent to ordinary percolation, details of which have been given by Stauffer and Aharony \cite{Stauffer1994}. The trapping version of the model appears to be much more complicated \cite{Knackstedt2002}. A comprehensive review of the invasion percolation literature has been given by Ebrahimi \citep{Ebrahimi2010}.

An interesting aspect of invasion percolation is the occurrence of ``bursts". The failure of a strong bond (large $s$) allows fluid to enter a region where there are weaker bonds (smaller $s$). The failure of these bonds is considered to be a burst. Roux and Guyon \cite{Roux1989} defined the size of a burst to be the number of smaller $s$ values that follow each failure in the sequences. Thus there can be bursts within bursts. Maslov\cite{Maslov1995} and Paczuski et al.\cite{Paczuski1996} have carried out extensive studies of these burst statistics. They found the frequency-size statistics of bursts to be well approximated by power laws. Roux and Wilkinson \cite{Roux1988} associated invasion percolation bursts with resistance jumps observed in laboratory studies of mercury injection into a porous medium.

The purpose of this paper is to reintroduce a form of invasion percolation that is applicable to fracking. Before doing this, a brief discussion of the physics of fracking is given. Oil and gas producing reservoirs are made up of sediments and organic material. As the thickness of deposited sediments increases, the temperature of the sediments increase and oil and/or gas are generated from the organic material. Oil is generated from organic material in the ``oil window" (temperatures of $60-120\,^{\circ}\unit{C}$, depths of $2-4\unit{km}$). Gas is generated from the organic material in the ``gas window" (temperatures of $100-200\,^{\circ}\unit{C}$, depths of $3-6\unit{km}$).

The current principal focus of fracking is the recovery of oil and gas from very old (200-300 million years) tight shale rocks. Shales are very fine grained sedimentary rocks. Oil and gas shales typically have grain diameters of about 4 microns and gas and/or oil filled porosity of $2-12\%$. Capillary forces associated with the fine structure reduce granular permeabilities to very low values. However, the chemical reactions associated with oil and gas generation produce high pressures in the oil and gas shales. One consequence of these high pressures is the generation of extensive sets of natural hydraulic fractures. In relatively young (2-30 million years) oil and gas shales these natural fractures generally produces high fracture permeabilities.

While natural hydraulic fractures generate high permeabilities in young shale reservoirs, these fractures are often sealed by natural processes in old shale reservoirs. One sealing process is deposition of calcite in the fractures. The existence of the sealed natural fracture sets plays an essential role in the effectiveness of fracking in the extraction of oil and gas from tight shales. The calcite fills the natural fractures impeding the flow of injected fluids. However, the bond between the calcite and the shale is relatively weak, allowing the natural fractures to be reopened.

In the past 10 to 15 years a new approach to fracking has been successful in extracting large quantities of tight shale oil and gas. An essential feature of the new approach is the use of ``slickwater" as the fracking fluid where additives reduce the viscosity of the water. Large volumes of water, typically 5,000-15,000 \unit{m^3}, are used in each frack. The water is driven into the preexisting natural fractures reactivating them \cite{Engelder2009}. Observed associated microseismicity indicates shear offsets on these fractures \cite{Maxwell2011}. The preexisting stress field focuses the seismicity in a plane perpendicular to the maximum principal compressive stress. The maximum principal compressive stress is generally vertical so that the seismicity is concentrated in a horizontal plane. We will associate the microseismicity observed in fracking with the bursts associated with invasion percolation. We will also associate the efficiency of fracking in extracting oil and gas to the self-similar structure of invasion percolation clusters.

\section{Our Model}
Our invasion percolation model is based on the model given by Wilkinson and Barsony \cite{Wilkinson1984}. These authors considered invasion percolation from a point source in two and three dimensions. The result was a single growing cluster of invaded sites. The scaling properties of this model were studied by De Arcangelis and Herrmann \cite{Herrmann1990a}. 

While our model may be applied to any lattice in any number of dimensions , we consider only a two-dimensional square lattice as illustrated in Figure \ref{fig:procedure}. The sites are connected by bonds. The sites are considered to be pore spaces on sealed natural fractures that the injected fluid will fill. The bonds are the sealed pathways for fluid flow between the pore spaces. The injected fluid flows through the natural fractures when the bonds are opened. If fluid injection is relatively slow, the injection is resisted primarily by capillary forces rather than viscous forces. Our invasion percolation model neglects pressure drops associated with viscous flow. In Figure 1, occupied sites are shown as solid circles and adjacent accessible sites are shown as dashed circles. The numbers in the circles identify the sites. Opened bonds (open fractures) between sites are shown as double solid lines and unopened bonds (sealed pathways) are shown as dashed lines.

\begingroup
\begin{figure} 
\centering
\subfigure[]
{
\includegraphics[]{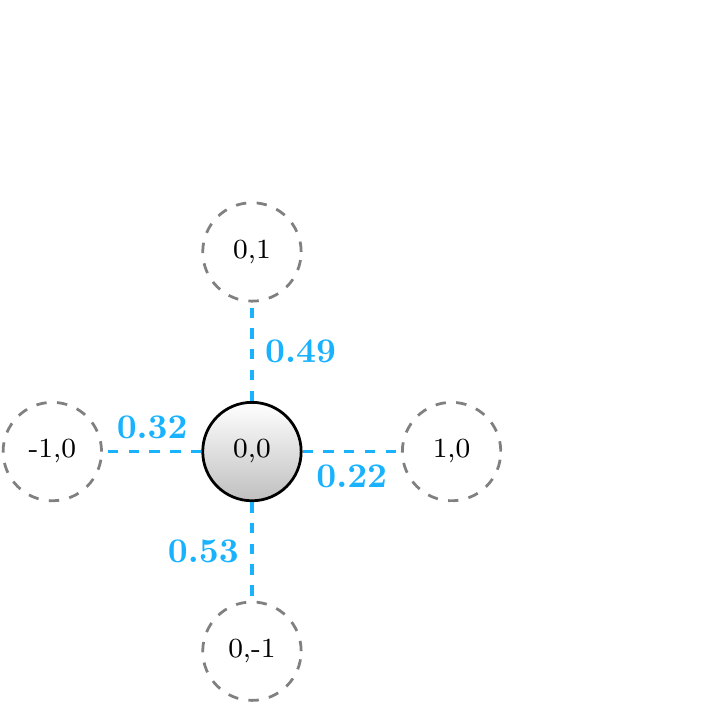}
\label{fig:origin}
}\qquad
\subfigure[]
{
\includegraphics[]{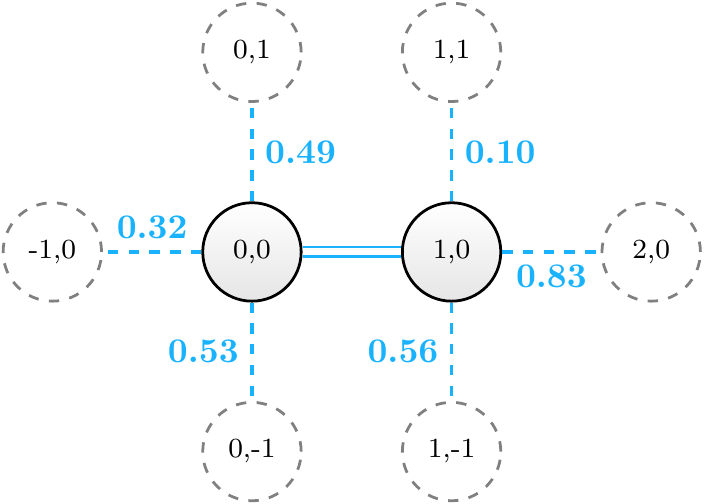}
}
\subfigure[]
{
\includegraphics[]{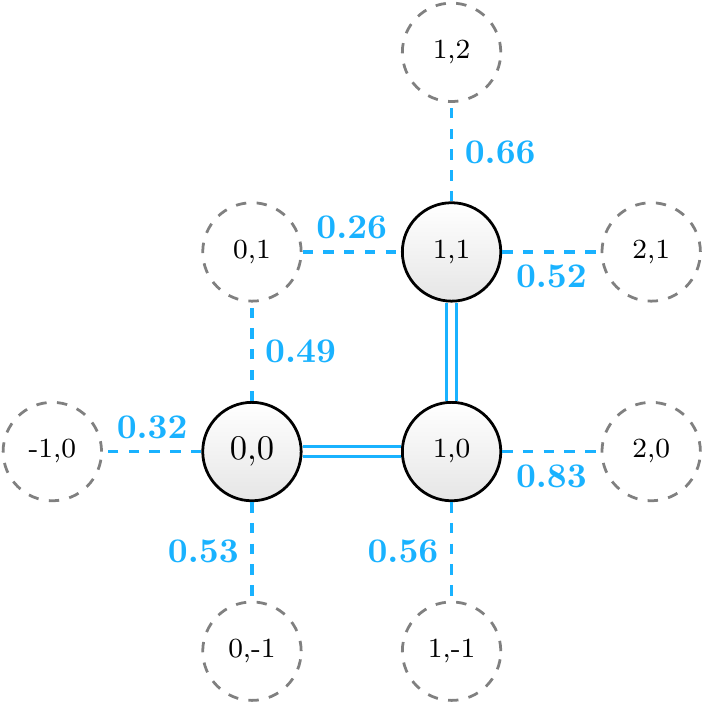}
}\qquad
\subfigure[]
{
\includegraphics[]{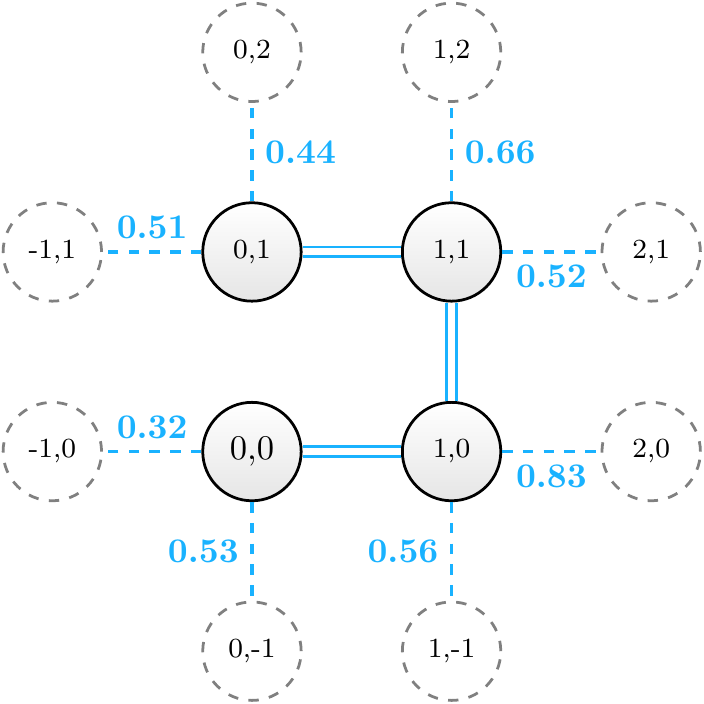}
\label{fig:trapped}
}
\caption{(Color online) Illustration of our invasion percolation model. Fluid filled sites are shown as solid circles and adjacent accessible sites are shown as dashed circles. The numbers in the circles identify the sites. Open bonds between fluid filled sites are shown as double solid lines. Unopened bonds between fluid filled and adjacent accessible empty sites are shown as dashed lines. The numbers next to the unbroken bonds are the random assigned strengths $s$. At each time step the weakest (smallest $s$) unbroken bond breaks and fluid flows to the adjacent accessible site. At any time there is an unbroken bond between two occupied sites, the bond is removed and remains closed for the rest of the simulation. A sequence of times steps is shown in Figures \ref{fig:origin} to \ref{fig:trapped}.}
\label{fig:procedure}
\end{figure}
\endgroup

High pressure fluid is introduced at the central site (0,0). The central cite is connected to the four neighboring sites (0,1), (1,0), (0,-1) and (-1,0) with closed bonds. These bonds are given random strengths $s$ in the range $[0,1]$. The random strengths $s=0.49$, $0.22$, $0.53$, $0.32$ of the bonds to the adjacent sites are shown in Figure \ref{fig:origin}. The weakest bond (smallest $s$) opens and the high pressure fluid flows into the adjacent site. In the example illustrated in Figure \ref{fig:procedure}(b) the weakest bond is (0,0)-(1,0) with $s=0.22$ and fluid fills site (1,0). This site is now connected with closed bonds to three neighboring sites (1,1), (2,0), (1,-1). These bonds are given random strengths $s=0.10$, $0.83$, $0.56$. On the next time step, illustrated in Figure \ref{fig:procedure}(b), fluid is allowed to flow through the weakest (smallest $s$) of the six available bonds, this bond is (1,0)-(1,1) with $s=0.10$. Site (1,1) is now connected with closed bonds to three neighboring sites (0,1), (1,2) and (2,1). These bonds are given random strengths $s=0.26$, $0.66$, $0.52$. On the next time step illustrated in Figure \ref{fig:procedure}(d), fluid is allowed to flow through the weakest (smallest $s$) of the eight available bonds, this bond is (0,1)-(1,1) with $s=0.26$ and site (0,1) is filled. Two new nearest neighbor bonds are added but the bond (0,0)-(0,1) is removed because both adjacent sites are filled.

We justify this removal because the two adjacent sites are at the same injection pressure so there is no differential pressure to open the bond. This bond removal condition is the major difference between our model and non-trapping invasion percolation. It prevents internal loops of opened bonds, thus the evolving cluster has a single path from the injection site to each filled site. This loop removal rule has been introduced previously by Barab\'{a}si \cite{Barabasi1996} for determining minimum energy spanning trees on a lattice and by Cieplak et al. \cite{Cieplak1996} to model optimal polymers in a strongly disordered medium. In the case of minimum energy spanning trees, it was shown that this model is simply the application of Prim's algorithm \cite{Prim1957} to a lattice. Also, we note that Sahimi et al. \cite{Sahimi1998} intoduced a more complicated invasion percolation model that also prevents the formation of internal loops. Additionally, a loopless version of regular percolation based upon the Leath algorithm \cite{Leath1976} has been introduced by Tzschichholz et al. \cite{Tzschichholz1989}.

\begin{figure}
\centering
\subfigure[]{\includegraphics[width=0.3\textwidth]{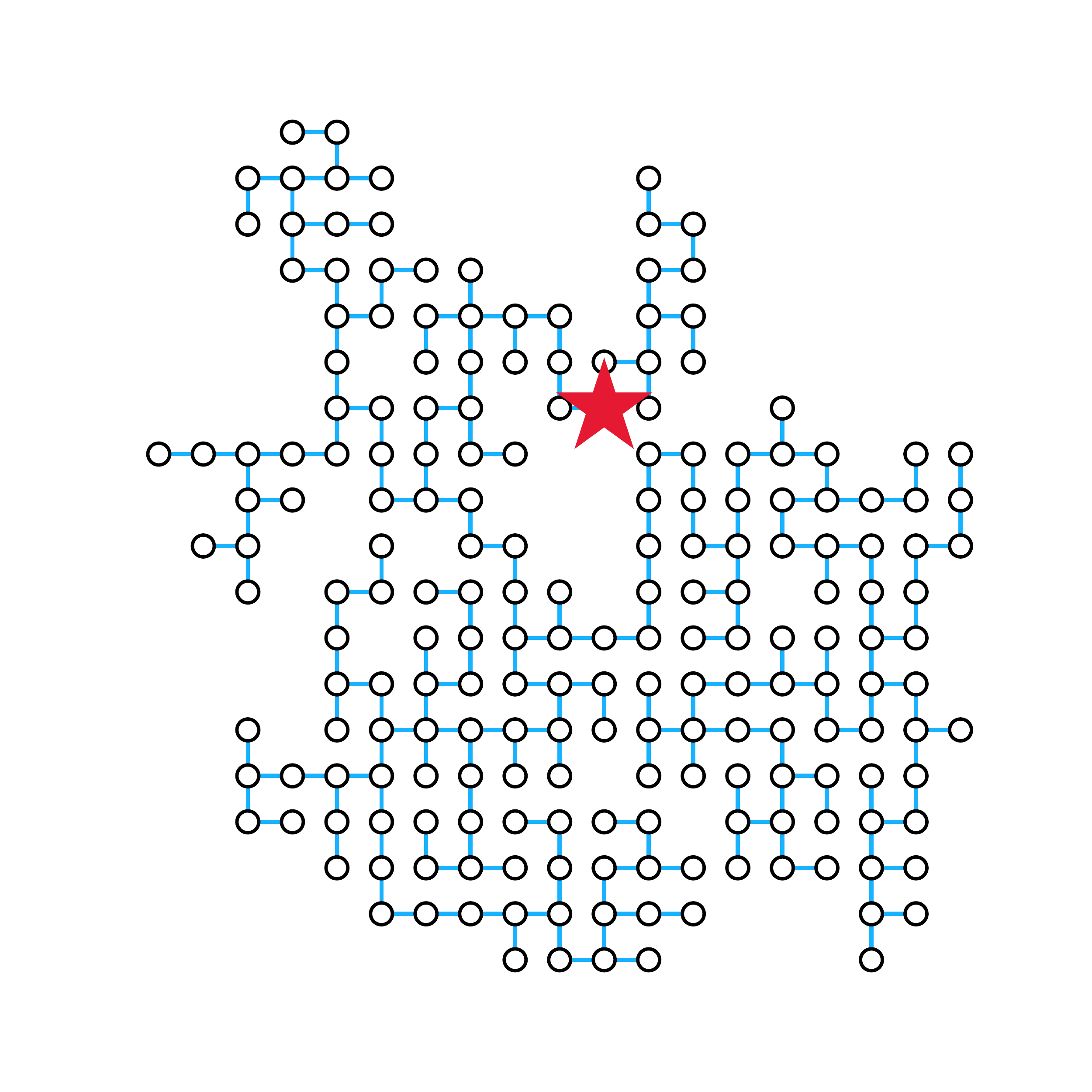}}
\subfigure[]{\includegraphics[width=0.3\textwidth]{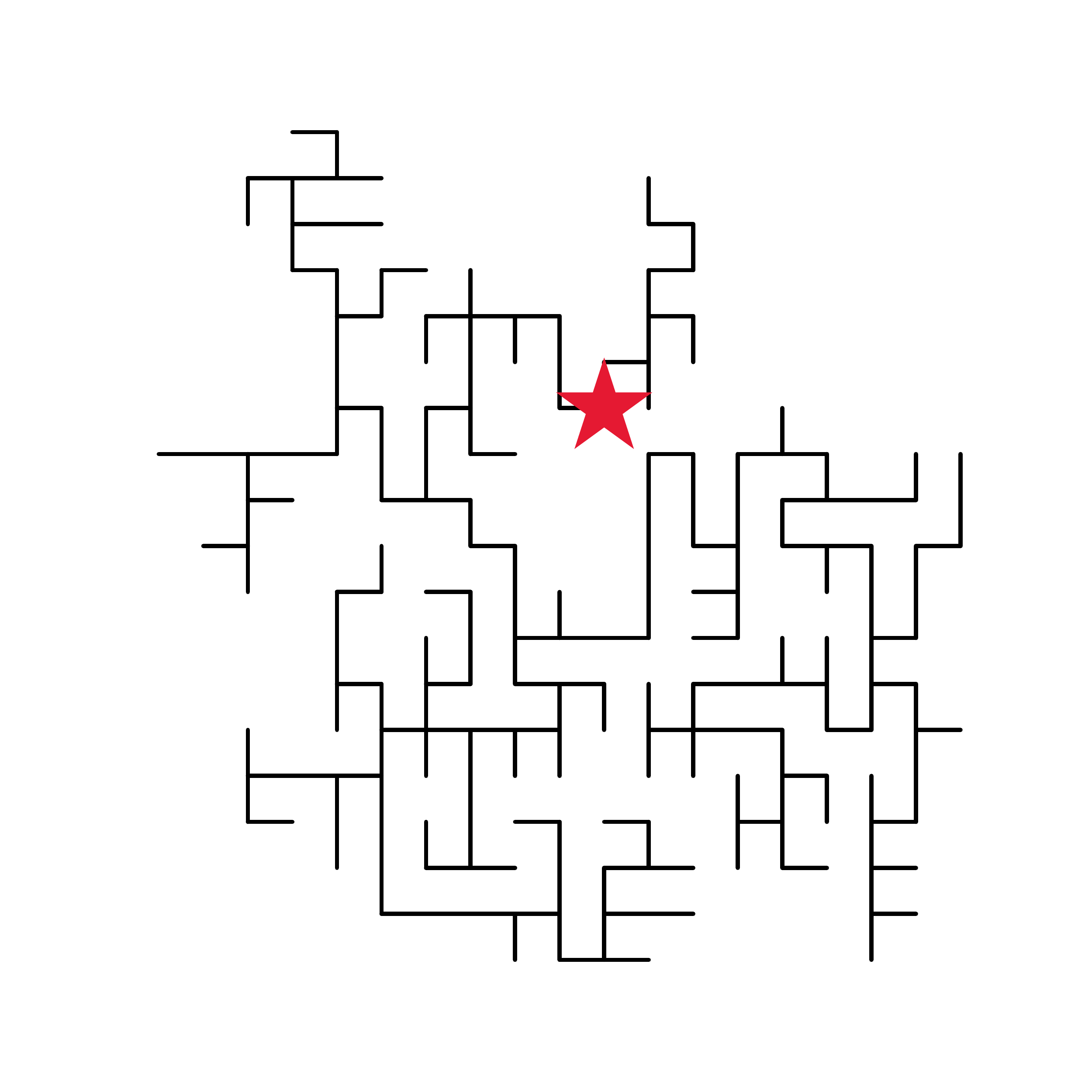}}
\subfigure[]{\includegraphics[width=0.3\textwidth]{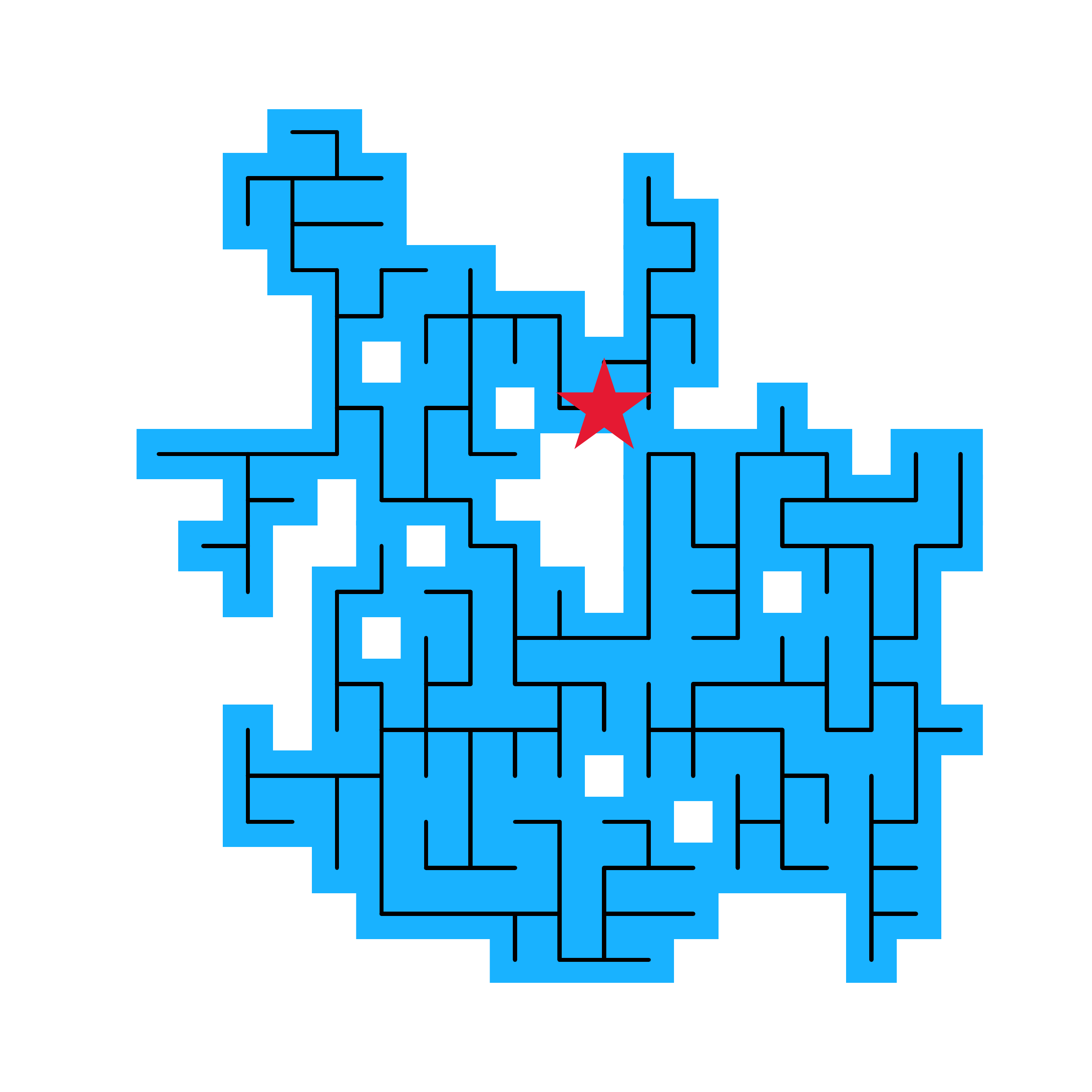}}
\caption{(Color online) Illustration of a small cluster with a mass $m=200$ (a) Fluid filled sites are shown as circles. The open bonds through which fluid flows are also shown. (b) Only the open bonds are shown. (c) Same as (a) except filled sites are shown as connected squares. The injection site is shown as the red star. There is a single path from each filled site to the injection site.}
\label{fig:cluster}
\end{figure}

The process illustrated in Figure \ref{fig:procedure} is continued forming a single connected cluster. At each time step, a bond is opened and an occupied site is added to the growing cluster. We will refer to the number of occupied sites in the growing cluster as the mass $m$ of the cluster. Because there is a one to one correspondence between an opened bond and a occupied site, $m$ is also the number of opened bonds in the cluster. The structure of our evolving cluster can be illustrated using either the filled sites or the opened bonds. A typical small cluster with mass $m=200$ is shown in Figure \ref{fig:cluster}. It can be clearly seen that the bond structure illustrated in Figure \ref{fig:cluster}(b) has no internal loops.

To prevent confusion, we wish to clearly define the terminology used in this paper for the various percolation models. Random percolation (RP) is the traditional, noninvasion model covered in detail by Stauffer and Aharony \cite{Stauffer1994}. Loopless random percolation (LRP) is the modification of this model introduced by Tzschichholz et al. \cite{Tzschichholz1989}. Non-trapping invasion percolation (NTIP) is a variant of invasion percolation where the defending fluid is compressible and invasion can occur along the entire cluster perimeter. Trapping invasion percolation (TIP) is the more common variant where the defeding fluid is incompressible and invasion can only occur along portions of the cluster perimeter that are connected to the edges of the lattice. Loopless non-trapping invasion percolation (LNTIP) is the invasion percolation model presented in this paper. To our knowledge there has not be a study of loopless trapping invasion percolation (LTIP). 

While our model has been introduced previously \cite{Barabasi1996, Cieplak1996}, there has not been very much work to quantify the model. In particular, the associated fractal dimensions of the growing cluster have not been determined. These values are necessary to test the hypothesis that this model (LNTIP), non-trapping invasion percolation (NTIP), random percolation (RP), and loopless random percolation (LRP) all belong to the same universality class \cite{Barabasi1996}. Additionally, the statistics of the bursts and the statistics of cluster structures have not been considered. These will be a focus of this paper.

\begin{figure}
\centering
\subfigure[$\;m=1000$]{\includegraphics[width=0.3\textwidth]{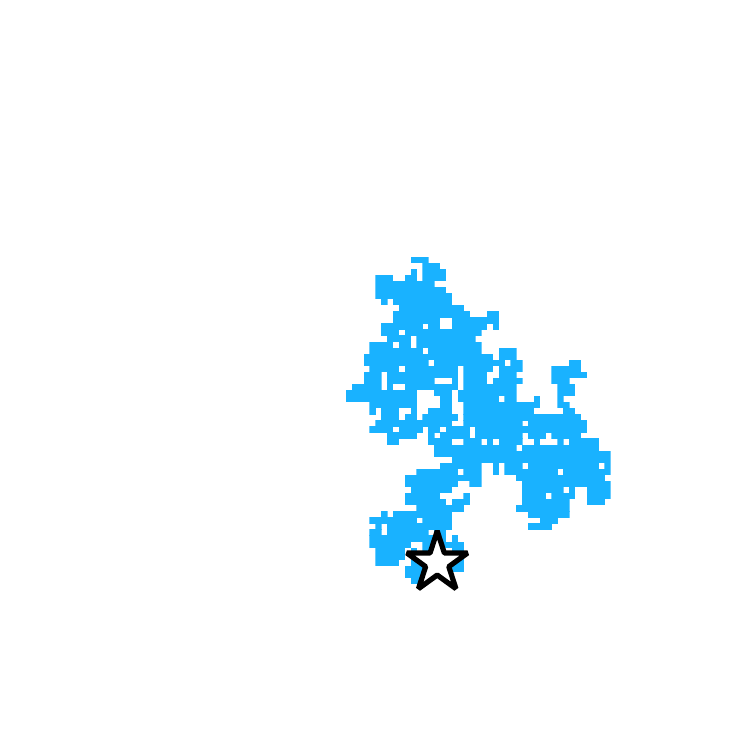}}
\subfigure[$\;m=2000$]{\includegraphics[width=0.3\textwidth]{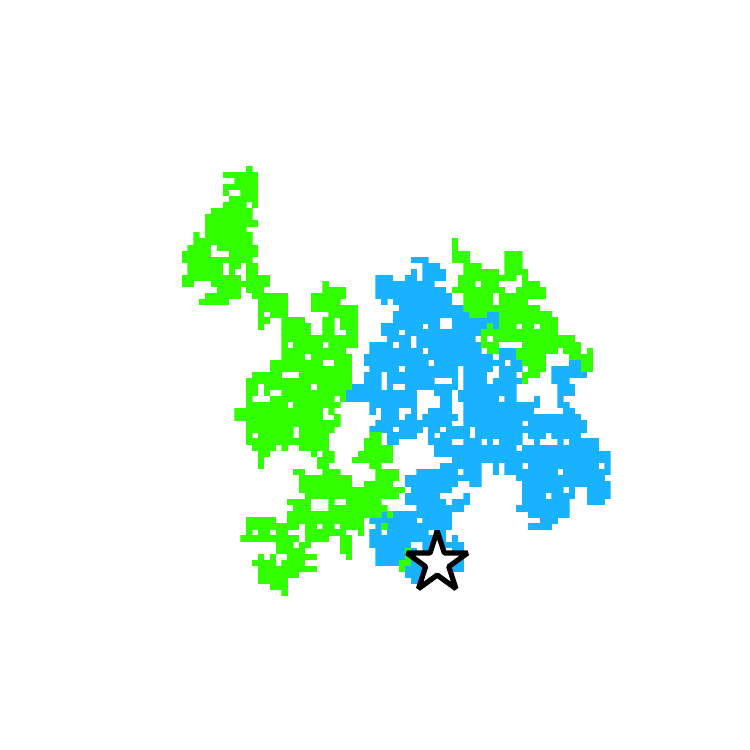}}
\subfigure[$\;m=3000$]{\includegraphics[width=0.3\textwidth]{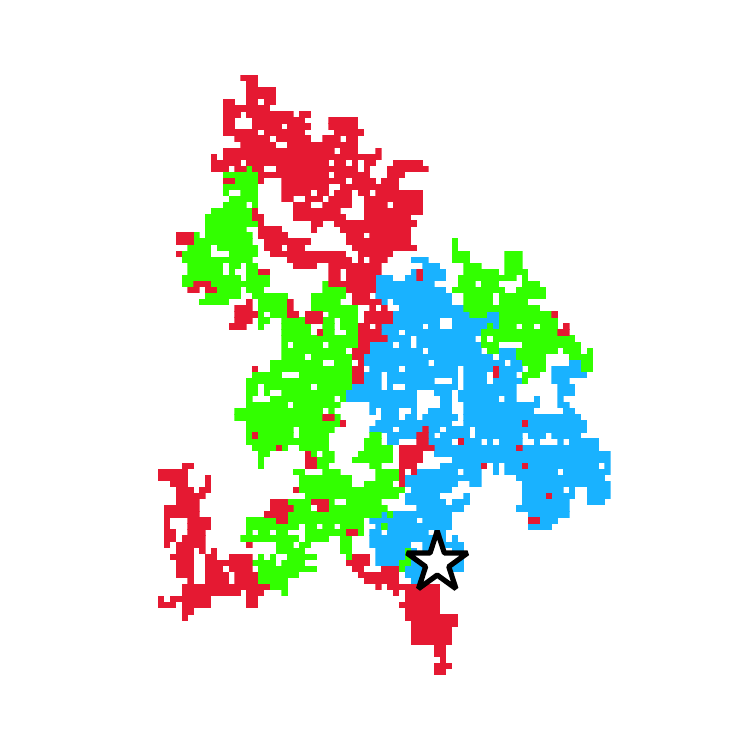}}
\caption{(Color online) Growth of a typical invasion percolation cluster. Three steps in the growth are shown. The first 1000 sites invaded are shown in blue (medium grey), the second 1000 in green (light grey), and the third 1000 are shown in red (dark grey). The injection site is shown as a black star.}
\label{fig:clusterGrowth}
\end{figure}

We have carried out extensive numerical simulations of our 2D invasion percolation model. The evolution of a typical cluster is shown in Figure \ref{fig:clusterGrowth}, three times during the growth of the cluster are shown. There are many interior regions with no occupied sites. The boundaries of these regions have relatively strong bonds (high $s$) which prevent further invasion but are not explicitly trapped. Within these cutoff regions, bonds with low values of $s$ are present, but they are not accessible. 

\section{Statistics of Grown Clusters}
\begin{figure}
\centering
\includegraphics[width=0.75\textwidth]{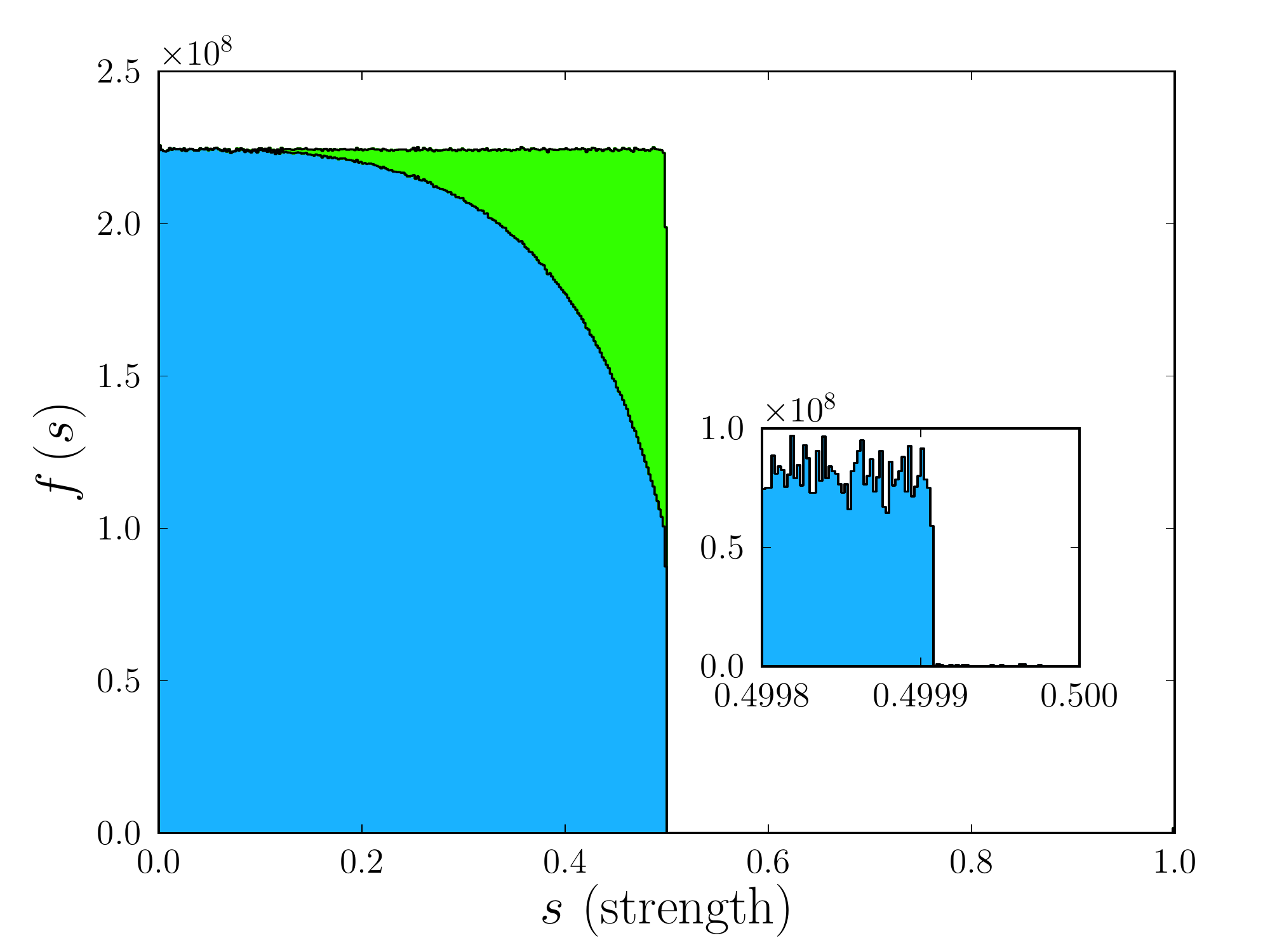}
\caption{(Color online) Frequency density of open bonds $f\left(s\right)$ is given as a function of bond strength $s$ in blue (dark gray). The data are from a cluster with 100 million occupied sites. The green (light grey) region adds the unopened bonds that have been removed (i.e. bond (0,0)-(0,1) in Figure \ref{fig:procedure}(d)). The inset gives the frequency density in the vicinity of the cutoff near $s=0.5$. The cutoff is very sharp at about $s=0.4999$ for our cluster of 100 million occupied sites (open bonds). The small number of strengths greater than this cutoff are part of an initial transient occurring while the cluster is smaller than $m=200,000$.}
\label{fig:hist}
\end{figure}

At each time step the weakest bond to a adjacent empty site breaks and the adjacent site fills with fluid. The invading cluster grows outward as illustrated in Figure \ref{fig:clusterGrowth}. It is clear from this figure that the boundary of the growing cluster is complex with fingers of occupied sites extending in all directions. It is of interest to study the distribution of strengths $s$ of the bonds that have opened. In Figure \ref{fig:hist} the frequency density of open bond strengths $f\left(s\right) = \frac{\mathrm{d}n}{\mathrm{d}s}$ is given as a function of $s$ in blue. The data are from a cluster with $m=100$ million occupied sites. The green region adds the strengths of unopened bonds that have been removed (i.e. bond (0,0)-(0,1) in Figure \ref{fig:procedure}(d)). The blue and green region is the frequency density of both the open and removed bonds. The rollover of open bond data is attributed to the systematic removal of relatively strong bonds. This data are characterized by the very strong cutoff. In order to study the nature of the strong cutoff, we give in Figure \ref{fig:hist} the frequency density of open bond strengths as a function of $s$ in the vicinity of the cutoff. The cutoff is very sharp at about $s=0.4999$ for this cluster of $m=100$ million occupied sites (open bonds). The small number of strengths greater than this cutoff are part of an initial transient occurring while the cluster is smaller than $m=200,000$.

As we have noted our problem is basically bond percolation as we utilize a statistical distribution of bond strengths in our model. It is important to note that the critical probability for the creation of a spanning cluster in bond percolation is $p_c=0.5$ \citep{Stauffer1994}. Our invasion percolation model also creates a spanning cluster so that it is not surprising that this requires the breaking of bonds weaker than $s=0.5$.

\begin{figure}
\centering
\subfigure[]{\includegraphics[width=0.375\textwidth]{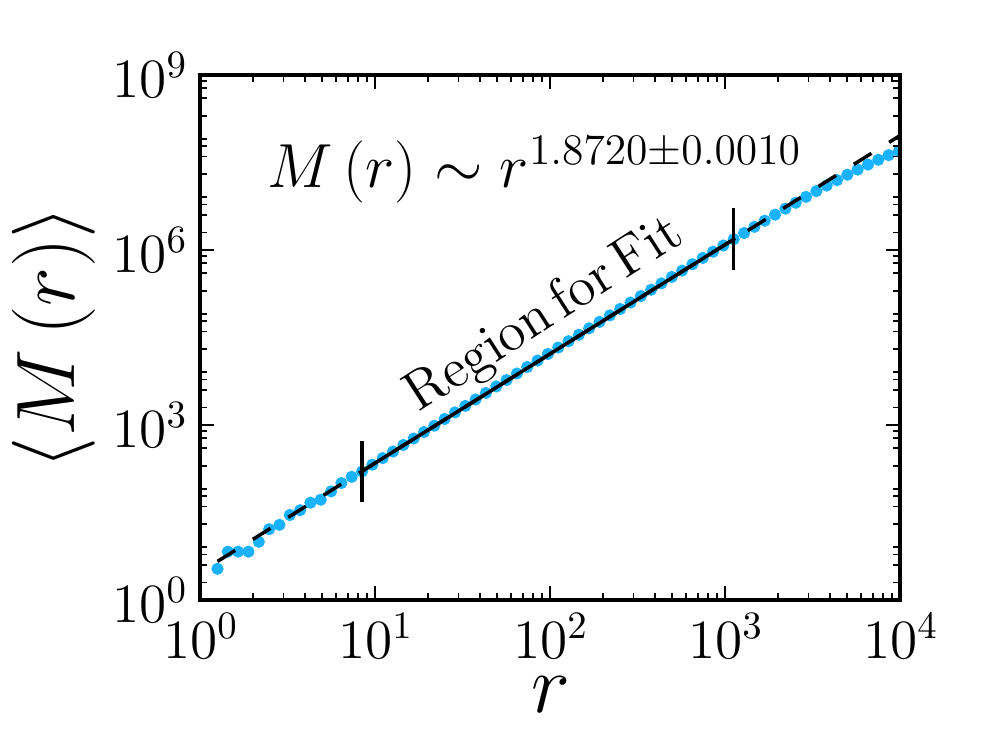}}
\subfigure[]{\includegraphics[width=0.375\textwidth]{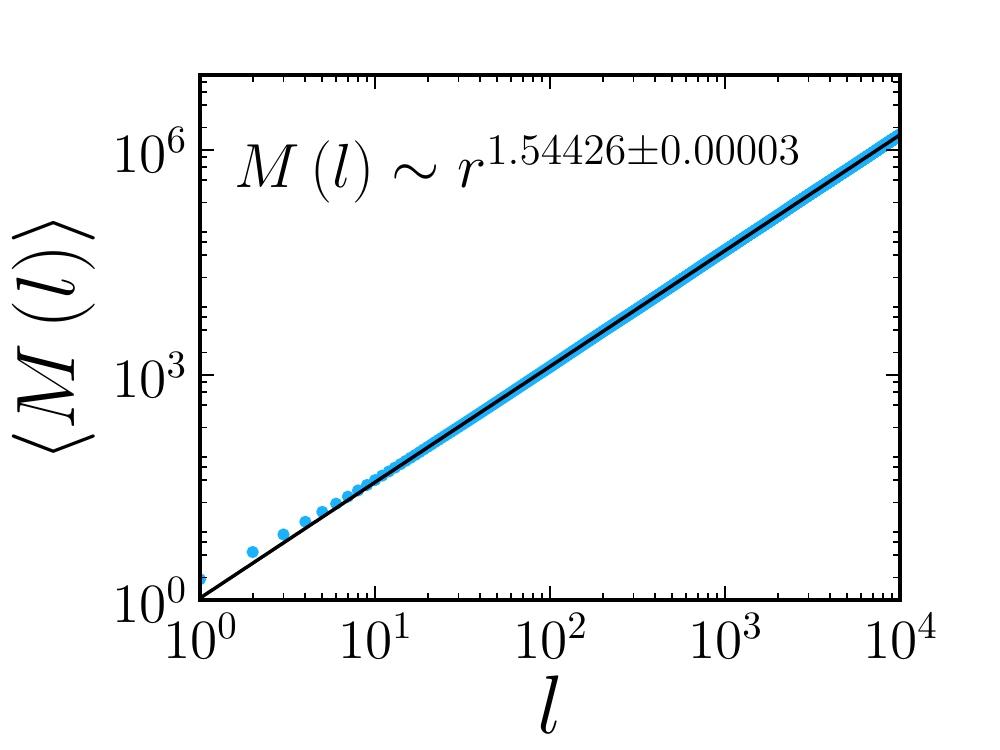}}
\caption{(Color online) (a) Dependence of the number of occupied sites $M$ contained within a circle of radius $r$ centered on the injection site on the radius $r$. The data are averages for 1000 realizations of clusters of mass $m=10^8$. The best fit of Eq. \eqref{eq:fractalDim} to the data gives a fractal dimension $d_f=1.8769$. (b) Dependence of the number of occupied sites $M$ with a chemical distance from the origin less than or equal to $l$ on the chemical distance $l$. The data are averages for 1000 realizations of clusters of mass $m=10^8$. The best fit of Eq. \eqref{eq:chemicalDim} to the data gives a chemical dimension $d_l=1.54628$.}
\label{fig:fractalDims}
\end{figure}

To characterize clusters grown using our model and to allow comparisons with other growth models, we utilize the approach given by Bunde and Havlin \cite{Bunde2012} and determine the fractal and chemical dimensions of clusters grown using our model. The fractal dimension $d_f$ is associated with a power-law dependence of the number of occupied sites $M$ contained within a circle of radius $r$ centered on the injection site on the radius $r$.
\begin{equation}
M\left(r\right) \sim r^{d_f}
\label{eq:fractalDim}
\end{equation}
We plot the number of occupied sites $M$ as a function of $r$ over several orders of magnitude. The fractal dimension of the cluster is the slope of the straight line through the data on a log-log plot. We have obtained the mass data as a function of $r$ for 1000 realizations with $m=100$ million occupied sites each. The average values obtained for these realizations are shown in Figure \ref{fig:fractalDims}a. For small values of $r$, the data contains artifacts that arise from the discrete nature of the lattice. For large values of $r$, nearly the entire cluster is contained within a radius $r$ and the slope flattens out as the radius approaches the cluster size and the cluster begins to appear point-like. We chose a region for determining a linear fit of the log-log data that excludes both of these artifacts. Using the aggregated log-log data we do a least squares fit to Eq. \eqref{eq:fractalDim} as shown in Figure \ref{fig:fractalDims}a. The fit gives a fractal dimension of $d_f=1.8769$. Because we performed the fit over an arbitrary region, the error shown in Figure \ref{fig:fractalDims}a represents the uncertainty in the fit and not necessarily the uncertainty in the fractal dimension.

We next turn out attention to the chemical dimension $d_l$ of our clusters. We define the chemical distance $l$ as the number of bonds between two sites along the cluster. The chemical dimension is of particular interest in polymer science and is associated with a power-law dependence of the number of occupied sites $M$ with a chemical distance from the origin less than or equal to $l$ on the chemical distance $l$
\begin{equation}
M \sim l^{d_{l}}
\label{eq:chemicalDim}
\end{equation}
We have also determined the chemical dimension of clusters grown using our model. We do so by keeping track of the chemical level of the bonds as they are added. Bonds connected to the origin are in the first chemical level ($l=1$). As bonds become available they belong to the next chemical level ($2$). The cluster is grown to a specified size $m$ and the number of bonds in each chemical level is determined. We plot the number of occupied sites $M$ as a function of $l$ over several orders of magnitude. The chemical dimension of the cluster is the slope of a straight line through the data on a log-log plot. The mass data as a function of chemical level $l$ were obtained for the same 1000 realizations used for the fractal dimension. The average values obtained for these realizations are shown in Figure \ref{fig:fractalDims}b. We did a least squares fit of the aggregated log-log data as shown in \ref{fig:fractalDims}b. This fit gives a chemical dimension of $d_l=1.54628$. We note that the error shown in Figure \ref{fig:fractalDims}b represents the uncertainty in the fit and not necessarily the uncertainty in the chemical dimension. We note that for small chemical levels, the average cluster size deviates from a power law due to the fixed coordination number of the square lattice.

Our demonstration of the validity Eqs. \eqref{eq:fractalDim} and \eqref{eq:chemicalDim} for our clusters implies a power-law scaling between $l$ and $r$ that defines a fractal dimension $d_{\mathrm{min}}$
\begin{equation}
l \sim r^{d_{\mathrm{min}}}
\label{eq:minDim}
\end{equation}
the three fractal dimensions discussed above are related by
\begin{equation}
M \sim \left(r^{d_{\mathrm{min}}}\right)^{d_{l}} \Rightarrow M \sim r^{ d_{\mathrm{min}} d_l} \Rightarrow d_{\mathrm{min}} = \frac{d_f}{d_l}
\label{eq:fractal_dim_relation}
\end{equation}
Taking $d_{f}=1.8769$ and $d_l=1.54628$ we find that $d_min=1.2138$.

We now compare the fractal dimension $d_f$, chemical dimension $d_l$, and $d_{\mathrm{min}}$ for our model (LNTIP) to the dimensions for random percolation (RP), loopless random percolation (LRP), site non-trapping invasion percolation (site NTIP), and bond trapping invasion percolation (bond TIP) on a square lattice. Because our model is inherently a growth model and because we make other comparisons later, we feel it appropriate to also compare our model to DLA. For RP in two dimensions $d_{f} = 91/48 \approx 1.896$, $d_l = 1.678 \pm 0.005$, and $d_{\mathrm{min}} = 1.13 \pm 0.004$ \cite{Bunde2012}.  For the LRP $d_{f} = 1.90 \pm 0.04$, $d_l = 1.68 \pm 0.02$, and $d_{\mathrm{min}} = 1.13 \pm 0.03$ \cite{Tzschichholz1989}. For site NTIP $d_{f} = 1.8959 \pm 0.0001$, $d_l = $, and $d_{\mathrm{min}} = 1.1307 \pm 0.0004$ \cite{Knackstedt2002}. We note that these values are consistent with the belief that NTIP is identical to RP. For bond TIP on a square lattice $d_{f} = 1.822 \pm 0.008$, $d_l = $, and $d_{\mathrm{min}} = 1.214 \pm 0.002$ \cite{Knackstedt2002}. As larger DLA clusters have been studied, it has been shown that DLA clusters on a square lattice are not self-similar \cite{Meakin1987}. However, we give results that are useful for comparison and are valid for small ($N < 10^4$) DLA clusters. For square lattice DLA simulations in two dimensions $d_{f} = 1.69 \pm 0.24$, $d_l = 1.69 \pm 0.05$, and $d_{\mathrm{min}} = 1.0 \pm 0.02$ \cite{Stanley1984}. These values are compared with our values given previously in Table \ref{table:fractal_dim_comparison}. 
\begin{table}
\setlength{\tabcolsep}{10pt}
\begin{tabular}{l c c c c c}
 & $d_{f}$ & $d_{l}$ & $d_{\mathrm{min}}$ \\
LNTIP (This Paper) & $1.8769$ & $1.54628$ & $1.2138$ \\
RP \cite{Bunde2012} & $91/48 \approx 1.896$ & $1.678 \pm 0.005$ & $1.13 \pm 0.004$ \\
LRP\cite{Tzschichholz1989} & $1.90 \pm 0.04$ & $1.68 \pm 0.02$ & $1.13 \pm 0.03$ \\
Site NTIP \cite{Knackstedt2002} & $1.8959 \pm 0.0001$ & $1.6767 \pm 0.0006$ & $1.1307 \pm 0.0004$ \\
Bond TIP (square lattice) \cite{Knackstedt2002} & $1.822 \pm 0.008$ & $1.5001 \pm 0.007$ & $1.214 \pm 0.002$ \\
DLA \cite{Stanley1984} & $1.69 \pm 0.24$ & $1.69 \pm 0.05$ & $1.0 \pm 0.02$
\end{tabular}
\caption{Comparison of three different fractal dimensions for loopless non-trapping invasion percolation (this paper), with random percolation (RP), loopless random percolation (LRP), site non-trapping invasion percolation (NTIP), bond trapping invasion percolation (TIP) on a square lattice, and DLA. The relationship between the fractal dimensions is given by Eq. \eqref{eq:fractal_dim_relation}.}
\label{table:fractal_dim_comparison}
\end{table}

The fractal dimension $d_f$ for LNTIP is slightly lower than RP, LRP, and NTIP. This is likely due to the systematic removal of bonds making our clusters less space-filling and lowering the fractal dimension. However, the fractal dimension $d_f$ for LNTIP is higher than that of bond TIP on a square lattice suggesting that the systematic removal of bonds in LNTIP is less extensive than the removal of bonds in TIP.  While all of the cluster models except TIP, have similar chemical dimensions ($\approx 1.68$), the chemical dimension of LNTIP is significantly lower leading to a higher $d_{\mathrm{min}}$.  This means that on average the shortest distance between two points along the cluster is longer for LNTIP  than for all other models considered, except in the case of TIP. In this case, the fractal dimention $d_{\mathrm{min}}$ for LNTIP is consistent with that of bond TIP. This is in agreement with the previous results of Porto et al. \cite{Porto1997}.
While the differences in fractal dimensions between LNTIP and the other models considered might make little difference in practice, LNTIP is significantly different from the other percolation models considered and is not part of the same universality class. The hypothesis that LNTIP, RP, LRP, and NTIP all belong to the same universality class is based on the assumption that the removal process is local and thus after a small renormalization, the removed bonds vanish. The assumption that the removed bonds vanish under renormalization is not true if the removed bonds are fractal and thus exist in some sense at all scales as was the case with loopless percolation \cite{Tzschichholz1989}. We find that the removed bonds in our model are fractal with a fractal dimension of $\approx1.877$ and thus never vanish after renormalization at any scale.

We argue that our model belongs to a distinct class from other percolations models; however, we do not claim that it belongs to a new universality class for the following reason. Universality in a very general sense is the notion that many systems have similar properties (scaling exponents, etc.) despite differences in the details of those systems (lattice type, etc.). A more restrictive, but very common requirement for a universality class is that the different systems possess the same set of scaling exponents. This definition is derived from the theory of renormalization where a universality class is associated with a single fixed point of the renormalization group and the critical exponents are associated with the relevant observables of that fixed point. A more complete description of the relationship between renormalization, critical exponents and universality can be found in \cite{Cardy1996}. The fact that LNTIP and TIP share a fractal dimension is evidence that our model is in fact nonuniversal. TIP has been shown to be nonuniversal with fractal dimensions that depend on the lattice type \cite{Knackstedt2002}. This suggests that our model might depend on the details of the lattice and thus our model is not universal. While our model belongs to a distinct class from other percolations models, more simulations using different lattice types are required to determine whether that class is universal.

\section{Bursts}
We now turn our attention to ``bursts" in our model. A burst is the breaking of a strong bond with a high strength $s$ followed by the breaking of a sequence of weak bonds with small strengths $s$. We associate ``bursts" in our model with the small seismic events that occur in fracking. Alternative definitions have been proposed for what constitutes a burst. Roux and Guyon \citep{Roux1988} defined a burst to start when $s_{i+1} < s_{i}$ in the sequence of breaking bonds. The burst sequence continued until $s_{i+n} > s_{i}$, the length of the burst is $n$. Maslov \citep{Maslov1995} and Paczuski et al. \citep{Paczuski1996} utilized this definition. This definition results in a hierarchy of bursts within bursts. Bursts have also been observed in more realistic invasion models \cite{Martys1991}.

\begin{figure}
\centering
\includegraphics[width=0.75\textwidth]{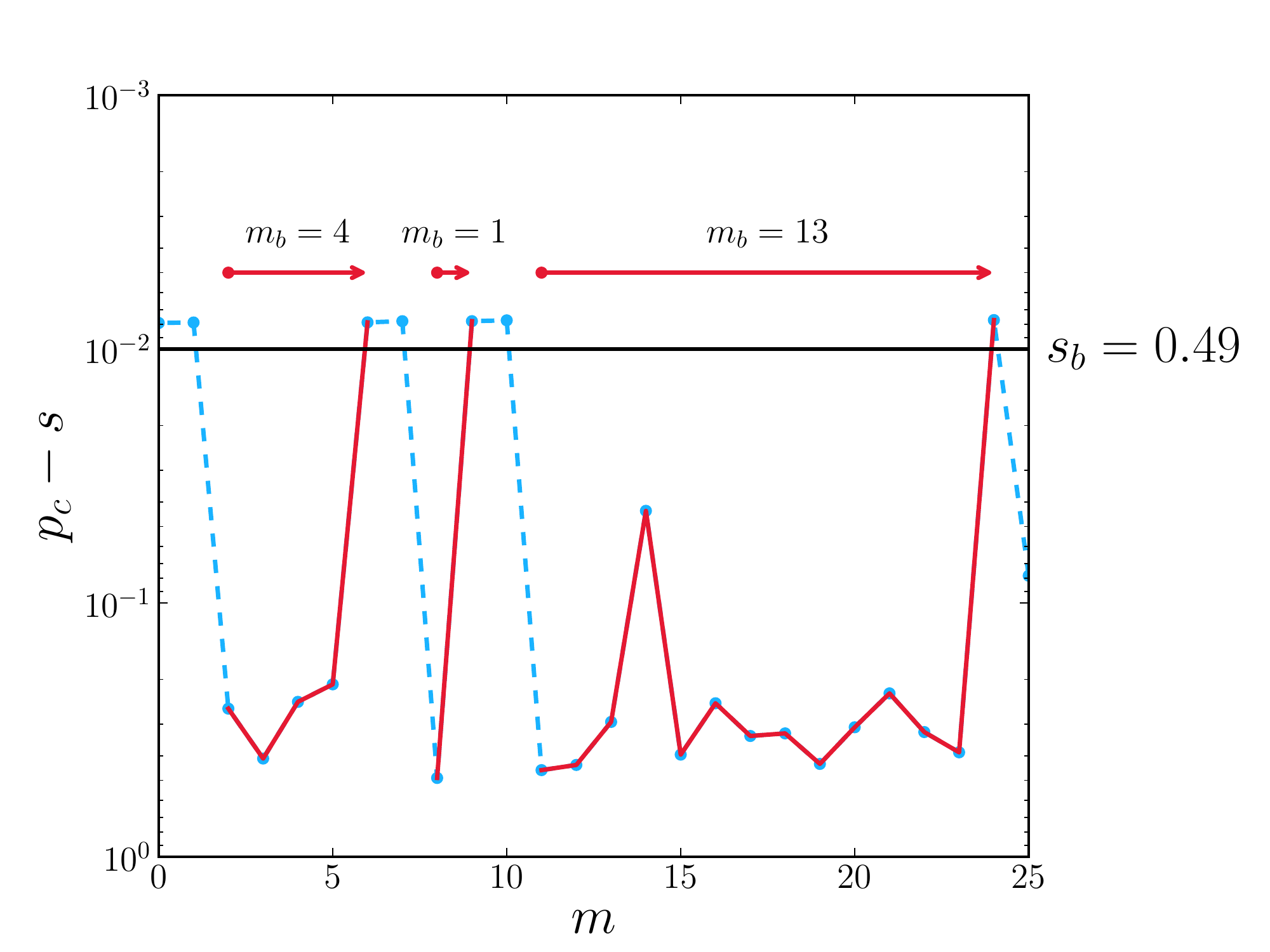}
\caption{(Color online) Illustration of our definition of a burst. A typical sequence of 25 opened bond strengths $s$ is given. The values of $p_c-s$ $\left(p_c=0.5\right)$ are shown as a function of time in dashed blue. We introduce a burst threshold $s_b=0.49$ $(p_c-s=0.01)$. A burst begins when an opened bond strength $s$ is smaller than $s_b$ and ends when an opened bond strength is greater than $s_b$. Three bursts with masses $m_b=4$, 1, and 13 are illustrated in solid red.}
\label{fig:timeSeries}
\end{figure}

In this paper, we propose an alternative definition of a burst that removes the hierarchical structure. Our definition is illustrated in Figure \ref{fig:timeSeries}. A typical sequence of 25 open bond strengths $s$ is given. This sequence was extracted after a run of 17,910 open bonds. The plot gives the values of $p_c-s$ with $p_c=0.5$. We introduce a burst threshold strength $s_b$. For the example given in Figure \ref{fig:timeSeries} we take $s_b=0.49$. A burst begins when an opened bond strength $s$ is smaller than $s_b$ and ends when an opened bond strength is higher than $s_b$. The bursts associated with the values of $s$ given in Figure \ref{fig:timeSeries} are shown. The number of opened bonds or filled sites in a burst is $m_b$. For the example illustrated $m_b=4$, 1, and 13.

\begin{figure}
\centering
\includegraphics[width=0.75\textwidth]{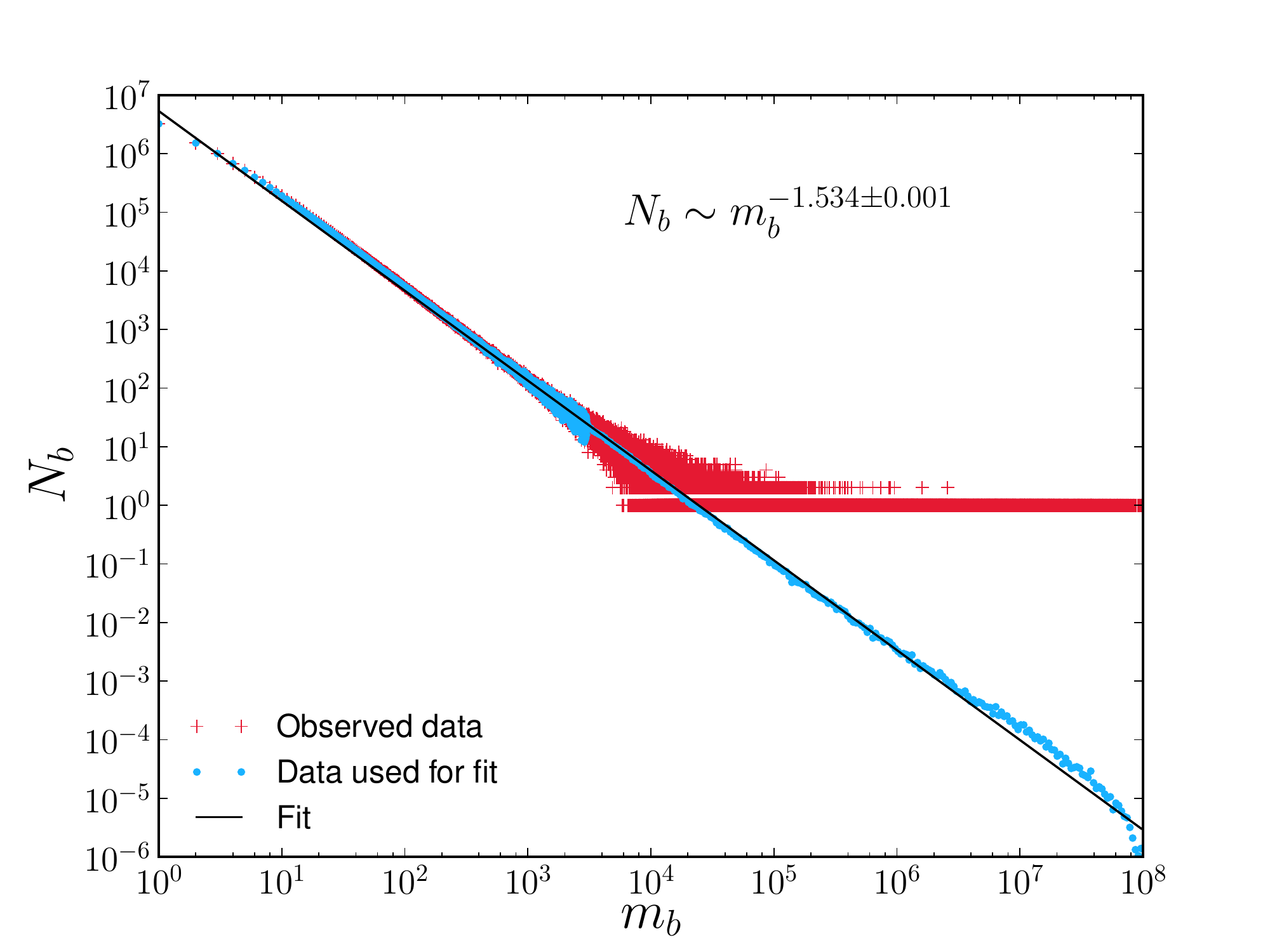}
\caption{(Color online) The dependence of the number of bursts $N_b$ with mass $m_b$ on $m_b$ aggregated from 1000 runs with $m=100$ million each. For $1 \le m_b \le 200$ we have $N_b > 10$ for all values of $m_b$, the data are shown as blue dots. For $m_b > 3000$ the observed data are shown as red crosses. The data are sparse with $N_b=0$ and 1 for many values of $m_B$. We bin this data and the binned data are shown as blue dots. The solid line is the least squared fit to the blue data as given in Eq. \eqref{eq:bursts}. The error estimate given is the error estimate of the fit. Because the fit depends on the choice of binning, the error estimate of the fit does not necessarily represent the uncertainty in the scaling exponent.}
\label{fig:burstDist}
\end{figure}

We now turn to the frequency size statistics of bursts. Our data are aggregated from 1000 realizations with 100 million time steps each ($m=t=1 \times 10^8$) and $s_b = 0.49995$. In Figure \ref{fig:burstDist} we give the aggregate number of bursts $N_b$ with mass $m_b$. For $1 \leq m_b \leq 3000$ we show the unit data (i.e $m_b=1,2,3,\dots$) in blue. For $m_b>2000$ we show this data in red. In this range the data are sparse and $N_b=0$ for many values of $m_b$. The standard treatment of this type of sparse data is to bin the available data \cite{Malamud2004}. Our binned data in this range are shown in blue. We have an excellent correlation of the blue data with the noncumulative distribution
\begin{equation}
N_b \sim m_b^{-1.534 \pm 0.001}
\label{eq:bursts}
\end{equation}
over seven orders of magnitude of cluster area $m_b$.

The cutoff value $s_b=0.49995$ that we have used is certainly arbitrary. However, we note that the value of cutoff used is very close to the critical probability for bond percolation on a square lattice $p_c = 0.5$. For smaller simulations ($m \approx 1$ million), we have determined the frequency-mass statistics for other cutoff values in the range $0.45 \le s_b \le 0.498$ and we find good power-law data in all cases with the same slope $\approx 1.53$. For comparison, the power-law slope in 2D non-trapping invasion peroclation was found to be $\approx1.60$ \cite{Maslov1995} and for the more realistic invasion model the slope was found to be $\approx1.35$ \cite{Martys1991}. It should be noted that in both cases the definition of a burst is different from what is presented in this paper.

We find that as simulations become larger and larger, a value of $s_b$ closer and closer to $p_c=0.5$ has to be used to produce good power-law data. We have also performed simulations on a triangular lattice (not given in this paper) and have found that using a cutoff close to the critical probability of $p_c = 0.34729$ also produces good power-law data. This suggests a relationship between the bursts and an underlying cluster structure of traditional percolation. In traditional percolation, the number of clusters of size $n_s$  is related to the size of a cluster $s$ near the critical point by the scaling relation
\begin{equation}
n_s \sim s^\tau
\label{eq:cluster_size_dist}
\end{equation}
The exponent $\tau$ is the Fisher exponent and $\tau = \frac{187}{91} \approx 2.05$ for traditional 2D percolation \cite{Stauffer1994}. We note the similarity between Eq. \eqref{eq:bursts} and Eq. \eqref{eq:cluster_size_dist}. Allthough there is a similarity between the distribution of static clusters in percolation and our bursts, the different growth procedure clearly leads to a different scaling exponent.

It has been noted that in finite lattices an effective critical occupation probability slightly less than the true critical occupation probability is required to obtain the proper critical point scalings \cite{Stauffer1994}. This suggests a reason why a cutoff just below the critical occupation probability is required to produce a power-law burst distribution. In the limit that simulations become very large, the cutoff required to produce a power-law burst distribution will become the critical occupation probability for traditional percolation.

\begin{figure}
\centering
\includegraphics[width=0.75\textwidth]{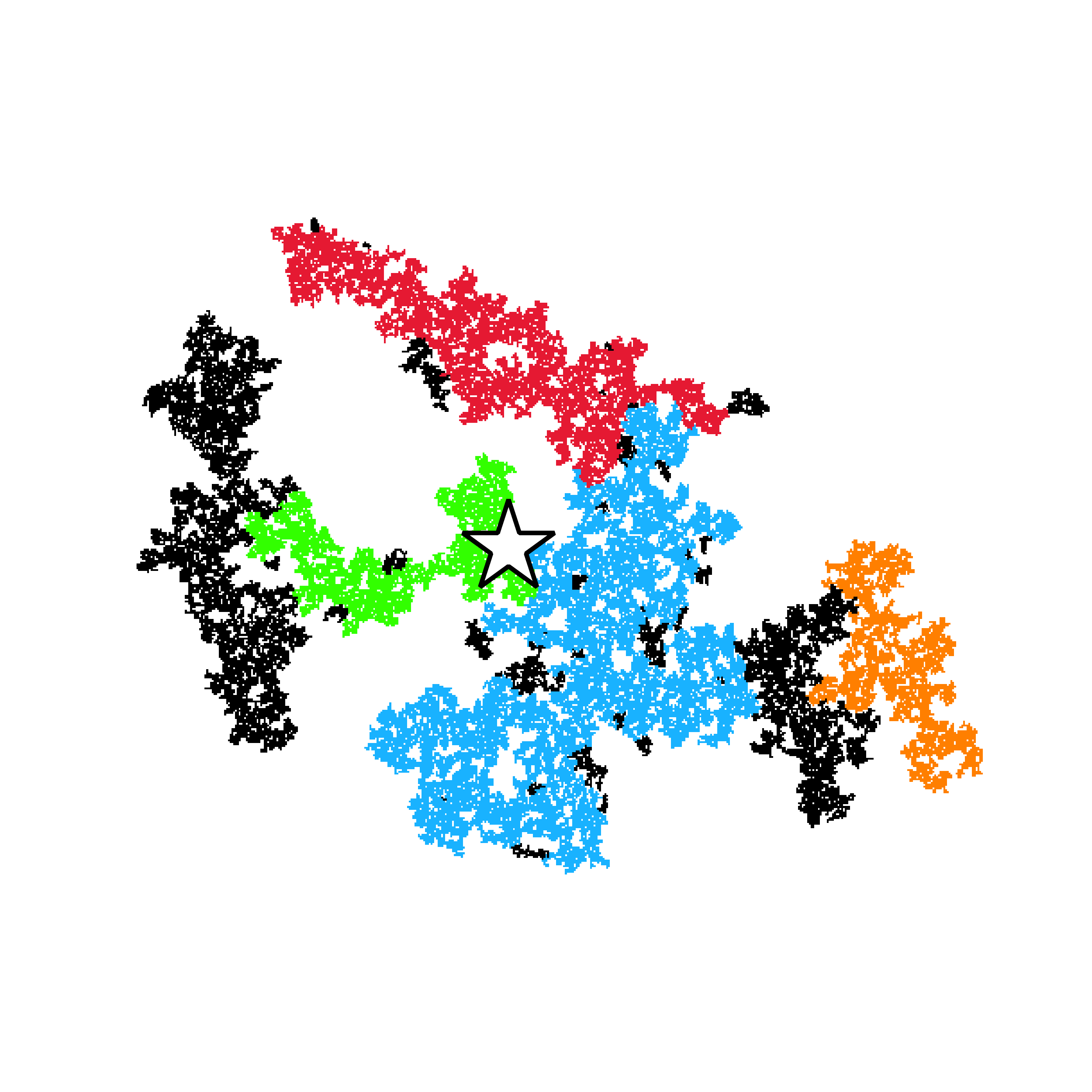}
\caption{(Color online) An example of burst structure for a simulation with mass $m=50,000$. The four largest bursts are shown in color. Smaller bursts and non-burst occupied sites are shown in black. The initiation site for the cluster is shown as a star.}
\label{fig:bursts}
\end{figure}

Bursts have spacial structures that resembles traditional percolation clusters. An example of burst structures is given in Figure \ref{fig:bursts}. The four largest bursts in a cluster with mass $m=50,000$ are shown in color. The smaller bursts and non-burst points are shown in black.

Gutenberg-Richter scaling for earthquakes gives a power-law dependence of the number earthquakes $N_e$ on the rupture areas greater than $A_r$ \cite{Turcotte1997}.
\begin{equation}
N_e = C A_{r}^{-b}
\label{eq:GRscaling}
\end{equation}
where $b$ has a universal value near unity. It is important to note that the data given in Fig. \ref{fig:burstDist} and the scaling given in Eq. \eqref{eq:bursts} are noncumulative whereas Eq. \eqref{eq:GRscaling} is cumulative. Maxwell \cite{Maxwell2011} determined the frequency magnitude scaling of the microseismic events associated with fracking and found that $b \approx 2$. More recent measurements of microseismicity generated during cold water injection into a geothermal reservoir found $b \approx 1.4$ \cite{Tafti2013}. The distribution of microseismicity associated with fracking as given by Maxwell \cite{Maxwell2011} and the observed power-law scaling indicates that the bursts statistics of our idealized invasion percolation model qualitatively represents the fluid migration in a fracking injection.

\section{Branching Statistics}
The network of fractures generated by a fracking injection is effective in extracting large quantities of oil and gas. In order to study the extraction associated with our invasion percolation model, we now consider the branching statistics of our clusters.

\begin{figure}
\centering
\subfigure[]{\includegraphics[width=0.49\textwidth]{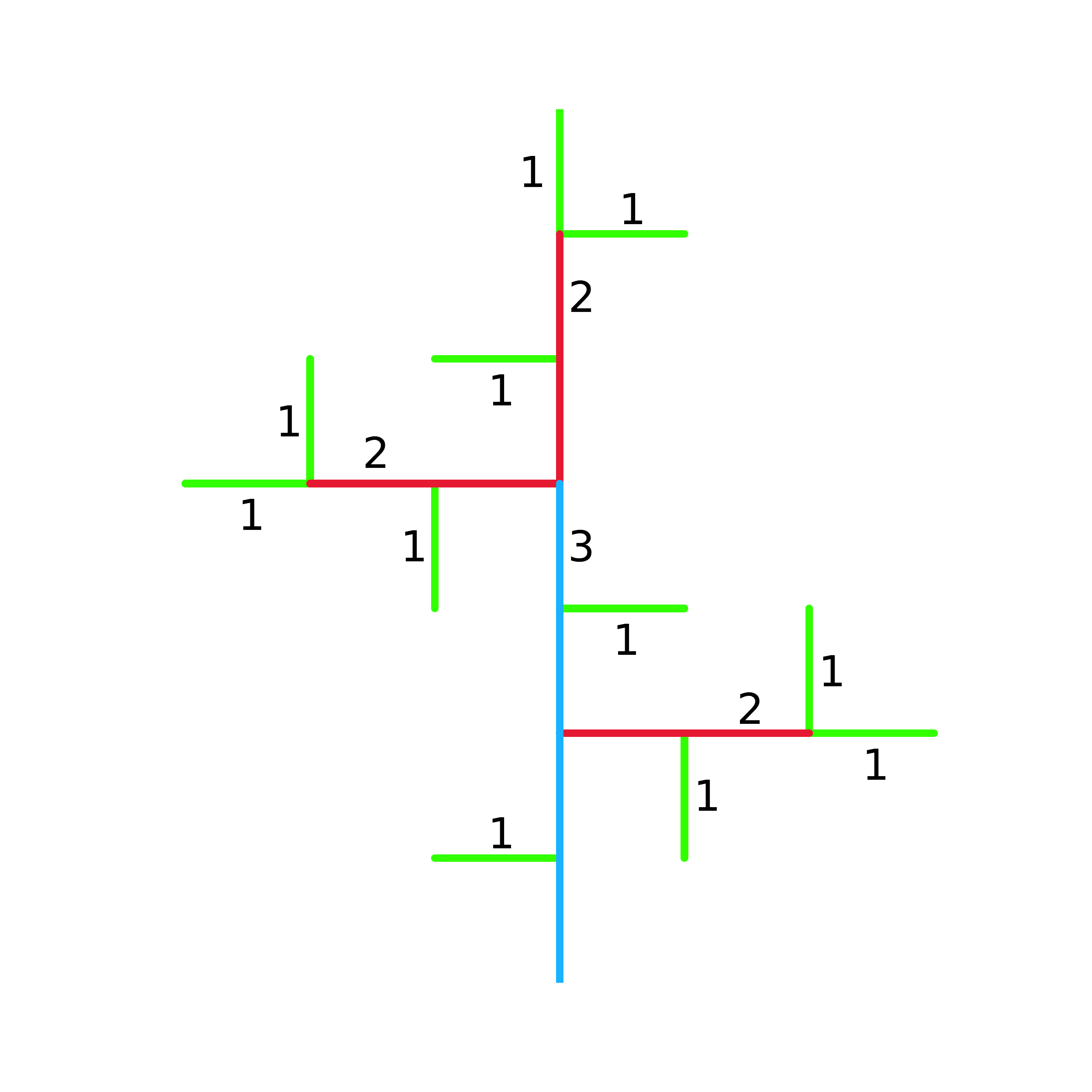}}
\subfigure[]{\includegraphics[width=0.49\textwidth]{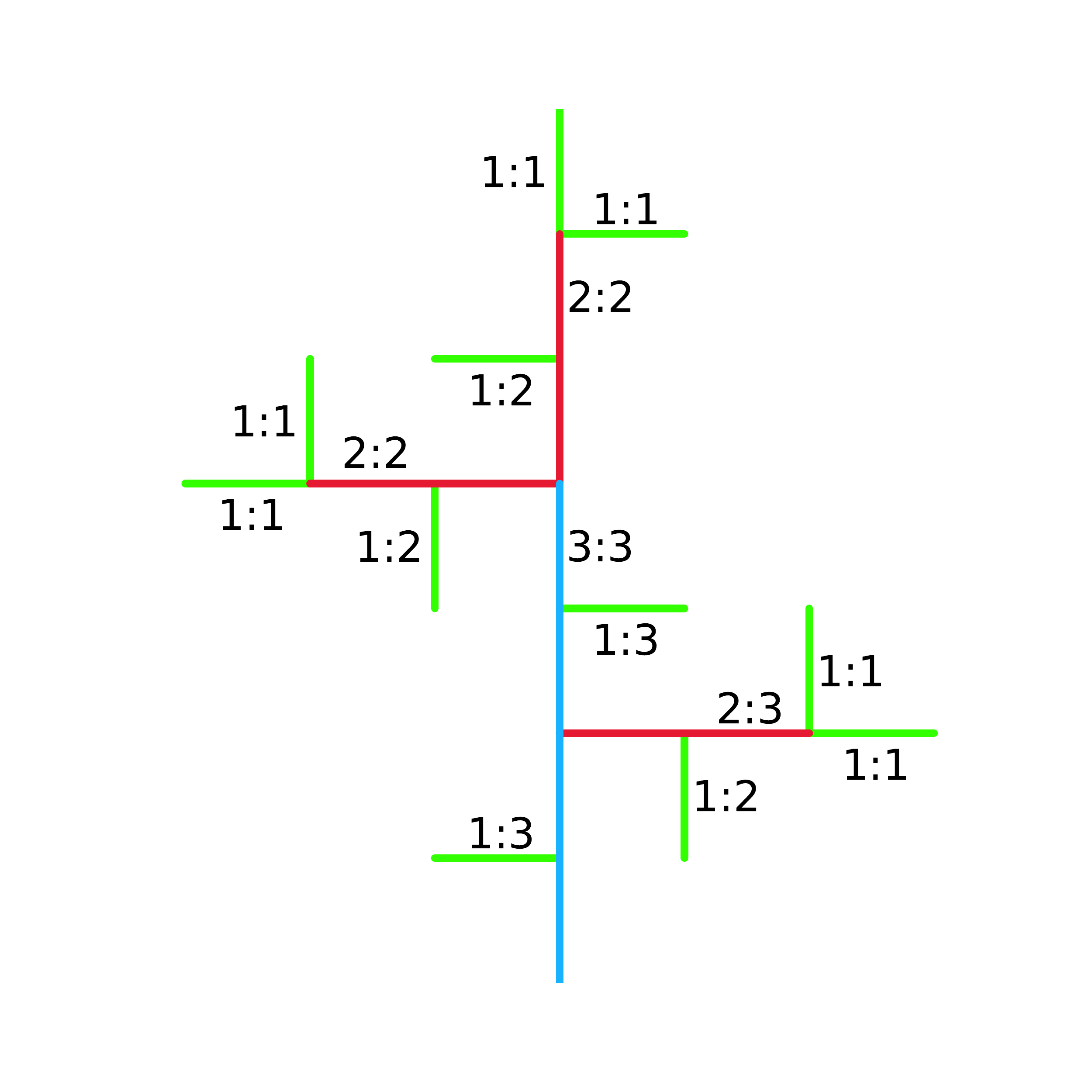}}
\caption{(Color online) Illustration of branch ordering statistics for a simple deterministic cluster. (a) Horton-Strahler definition of branch ordering. (b) Tokunaga definition of branch ordering taking account of side branching.}
\label{fig:branchingIllustration}
\end{figure}

Because the growing bond cluster forms a tree graph (contains no internal loops), it can be analyzed using the branch ordering statistics introduced by Horton \cite{Horton1945} and Strahler \cite{Strahler1952} for river networks. This ordering is illustrated in Figure \ref{fig:branchingIllustration}a. Tip branches (bonds) are defined to be first order $(i=1)$. When two first-order branches combine, they form a second order branch $(i=2)$, and so forth. The bifurcation ratio $R_b$ is defined by 
\begin{equation}
R_b = \frac{N_i}{N_{i+1}}
\label{eq:bifurcationR}
\end{equation}
where $N_i$ is the number of branches of order $i$. The length-order ratio $R_r$ is defined by
\begin{equation}
R_r = \frac{r_{i+1}}{r_i}
\label{eq:lengthR}
\end{equation}
where $r_i$ is the mean length of branches of order $i$. For a self-similar branching cluster the $R_b$ and $R_r$ are constant independent of $i$. In this case the fractal dimension of the cluster is given by
\begin{equation}
D = \frac{\ln{R_b}}{\ln{R_r}}
\label{eq:TfractalDim}
\end{equation}
Many natural phenomena are well approximated by self-similar branching statistics \cite{Turcotte1997}, one example is river networks.

\begin{figure}
\centering
\includegraphics[width=0.75\textwidth]{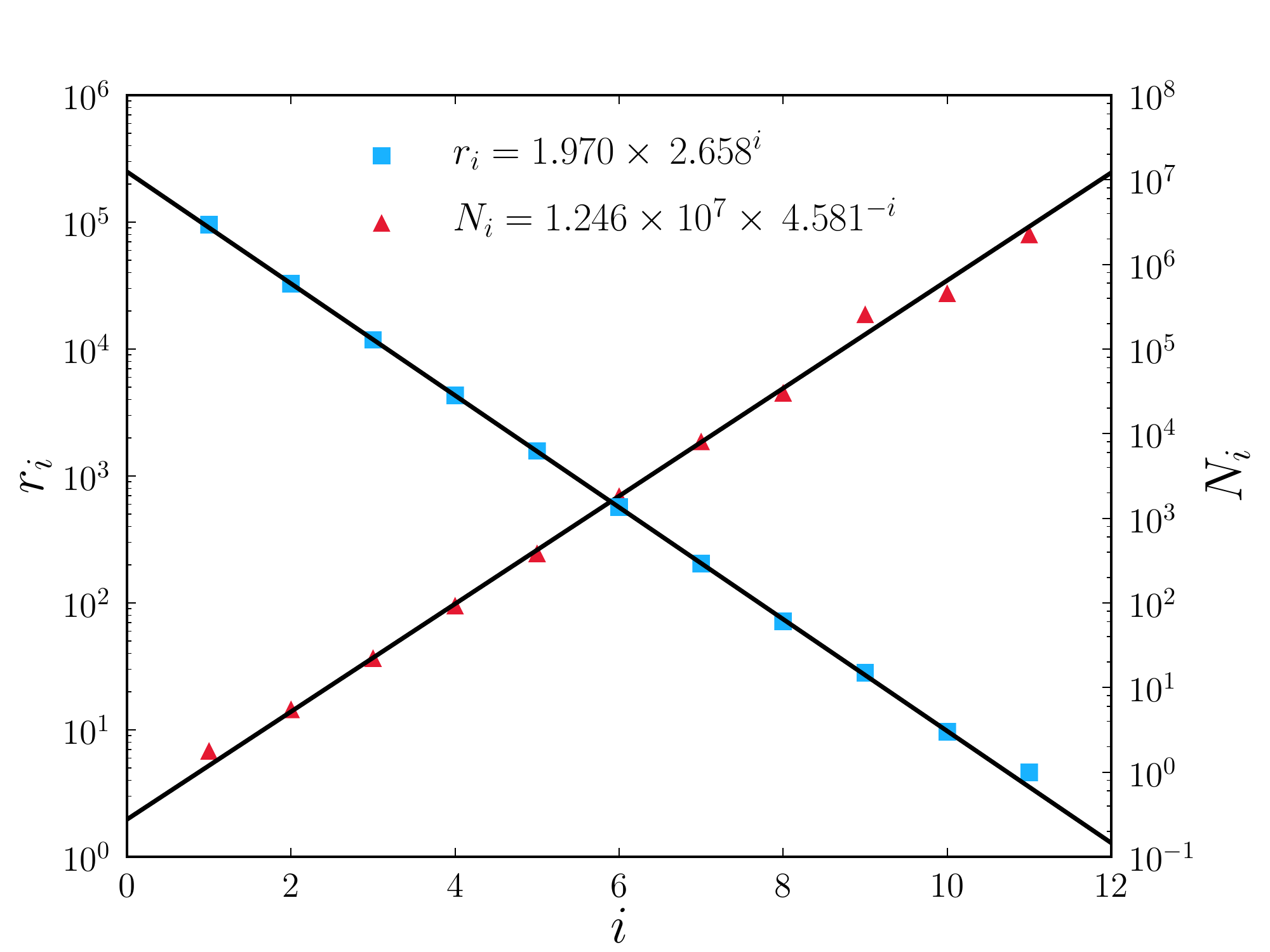}
\caption{(Color online) As blue squares, the mean length $r_i$ of branches of order $i$ is given as a function of $i$ for a cluster of mass 10 million. The blue squares are the data and the line is the best least squares fit to a linear correlation, from Eq. \eqref{eq:lengthR} we have a length-order ratio $R_r=2.658$. As red triangles, the number $N_i$ of branches of order $i$ is given as a function of $i$ for a cluster of mass 10 million. The red triangles are the data and the line is the least squares fit to a linear correlation, from Eq \eqref{eq:bifurcationR} we have the branching ratio $R_b = 4.581$.}
\label{fig:orders}
\end{figure}

We have obtained the Horton-Strahler branching statistics for a typical numerical simulation of our invasion percolation model. The branch-order statistics for a $m=10$ million cluster are given in Figure \ref{fig:orders}. This is a $n=11$ order cluster and the $N_i$ are given as a function of $i$. An excellent correlation with 
\begin{equation}
N_i = 1.246 \times 10^7 \times 4.581^{-i}
\end{equation}
is found. Thus the bifurcation ratio is nearly constant with a value $R_b = 4.581$. The length-order statistics for this cluster are given in Figure \ref{fig:orders}, the mean lengths $r_i$ of branches of order $i$ are given as a function of $i$. An excellent correlation with
\begin{equation}
r_i = 1.970 \times 2.658^i
\end{equation}
is found. Thus the length-order ratio is nearly constant with a value $R_r = 2.658$. Our invasion percolation cluster exhibits fractal behavior and the fractal dimension from Eq. \eqref{eq:TfractalDim} is $D=1.557$

An improved branch-ordering classification was introduced by Tokunaga \citep{Tokunaga1978, Tokunaga1984}. This ordering takes into account side branching and is illustrated in Figure \ref{fig:branchingIllustration}b. A first-order branch that joins another first-order branch is denoted $1:1$, a first-order branch that joins a second-order branch is denoted $1:2$, and so forth; $N_{ij}$ is the number of branches of order $i$ that join branches of order $j$. The total number of branches of order $i$, $N_i$, is related to the $N_{ij}$ by
\begin{equation}
N_i = \sum_{i=1}^n N_{ij}
\end{equation}
where $n$ is the branch order of the cluster.

The branch numbers $N_{ij}$ constitute a square upper triangular matrix. The $n=3$ branch-number matrix for the cluster illustrated in Figure \ref{fig:branchingIllustration}b is given in Table \ref{table:branchMatrix}a.
\begin{table}
\setlength{\tabcolsep}{8pt}
\begin{tabular}{c c c}
\begin{tabular}{r r r | r}
	$N_{1:1}=6$ & $N_{1:2}=3$	& $N_{1:3}=2$ & $N_1 = 11$	\\
	  & $N_{2:2}=2$	& $N_{2:3}=1$ & $N_2 = 3$	\\
	  & 	& $N_{3:3}=1$ & $N_3 = 1$	\\
\end{tabular}
& \hspace{0.5in} &
\begin{tabular}{r r r}
	$T_{1:2}=1$ & $T_{1:3}=2$ \\
	  & $T_{2:3}=1$ \\
\end{tabular}\\
(a) & \hspace{0.5in} & (b) \\
\end{tabular}
\caption{(a) Branch-number matrix and (b) the branching ratio matrix for the cluster illustrated in Figure \ref{fig:branchingIllustration}b.}
\label{table:branchMatrix}
\end{table}
The structure of branching clusters can be further classified in terms of branching ratios $Tij$. These are the average number of branches of order $i$ that join a branch of order $j$ and are defined by
\begin{equation}
T_{ij} = \frac{N_{ij}}{N_{j}}
\end{equation}
Again the branching ratios define a square, upper triangular matrix. The branching ratio matrix for the cluster illustrated in Figure \ref{fig:branchingIllustration}b is given in Table \ref{table:branchMatrix}b. Tokunaga \citep{Tokunaga1978, Tokunaga1984} defined a restricted class of self-similar branching networks by introducing the branching ratio $T_k = T_{i,i+k}$ and requiring that
\begin{equation}
T_k = a c^{k-1}
\label{eq:tokunaga}
\end{equation}
where a and c are constants. The example given in Figure \ref{fig:branchingIllustration}b and Table \ref{table:branchMatrix}b satisfies this condition since $T_1=1$, $T_2=2$, $a=1$, and $c=2$. This class of branching networks is known as Tokunaga networks.

\begin{figure}
\centering
\includegraphics[width=0.75\textwidth]{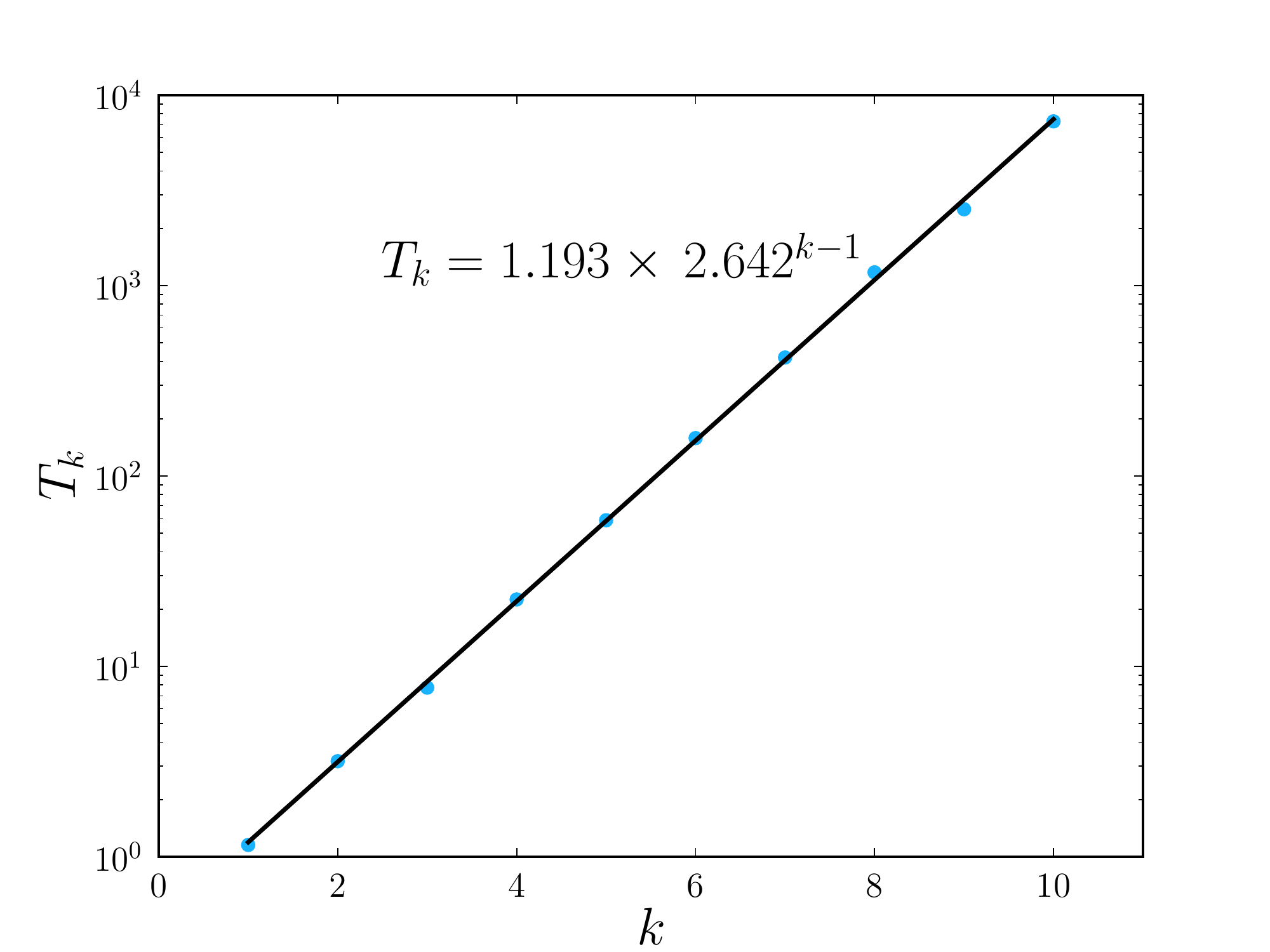}
\caption{(Color online) Dependence of the mean branching ratios $T_k$ for a cluster of mass $m=10$ million. The blue dots are the data and the black line is the fit.}
\label{fig:tokunaga}
\end{figure}

We now give the Tokunaga branching statistics for the mass $m=10$ million cluster considered above. The branch-number matrix is given in Table \ref{table:branchNum}. Values of $N_{ij}$ are given for this $n=11$ network as well as values of $N_{i}$. The branching-ratio matrix is given in Table \ref{table:branchRatio}, values of $T_{ij}$ are given. In order to test whether this is a Tokunaga network we determine mean values of $T_k$ using the relation
\begin{equation}
T_{k} = \frac{1}{n-k} \sum_{j=1}^{n-k} T_{i,j+k}
\end{equation}
The dependence of the mean branching ratios $T_k$ on $k$ is given in Figure \ref{fig:tokunaga}. An excellent correlation with Eq. \eqref{eq:tokunaga} is found taking $a=1.193$ and $c=2.642$, our invasion percolation cluster is a Tokunaga network to a good approximation.
\begin{table}
\scriptsize
\setlength{\tabcolsep}{5pt}
\begin{tabular}{c r r r r r r r r r r r | r}
 & \multicolumn{1}{c}{j=1} & \multicolumn{1}{c}{2} & \multicolumn{1}{c}{3} & \multicolumn{1}{c}{4} & \multicolumn{1}{c}{5} & \multicolumn{1}{c}{6} & \multicolumn{1}{c}{7} & \multicolumn{1}{c}{8} & \multicolumn{1}{c}{9} &\multicolumn{1}{c}{10} & \multicolumn{1}{c}{11} & \multicolumn{1}{c}{$N_i$} \\
i=1 & 1,190,268 & 767,355 & 417,458 & 243,197 & 139,578 & 85,601 & 49,716 & 24,880 & 25,414 & 7,441 & 7,287 & 2,958,195 \\  
2 & 0 & 257,348 & 139,793 & 81,452 & 47,524 & 29,333 & 16,963 & 8,627 & 8,878 & 2,662 & 2,554 & 595,134 \\
3 & 0 & 0 & 56,716 & 29,436 & 17,124 & 10,799 & 6,253 & 3,178 & 3,296 & 932 & 940 & 128674 \\
4 & 0 & 0 & 0 & 12,519 & 6,224 & 3,997 & 2,421 & 1,246 & 1,223 & 362 & 366 & 28,358 \\
5 & 0 & 0 & 0 & 0 & 2,710 & 1,438 & 886 & 428 & 502 & 157 & 138 & 6,259 \\
6 & 0 & 0 & 0 & 0 & 0 & 582 & 336 & 172 & 176 & 45 & 44 & 1,355 \\
7 & 0 & 0 & 0 & 0 & 0 & 0 & 122 & 60 & 69 & 71 & 23 & 291 \\
8 & 0 & 0 & 0 & 0 & 0 & 0 & 0 & 30 & 19 & 7 & 5 & 61 \\
9 & 0 & 0 & 0 & 0 & 0 & 0 & 0 & 0 & 6 & 5 & 4 & 15 \\
10 & 0 & 0 & 0 & 0 & 0 & 0 & 0 & 0 & 0 & 2 & 1 & 3 \\
11 & 0 & 0 & 0 & 0 & 0 & 0 & 0 & 0 & 0 & 0 & 1 & 1 \\
\end{tabular}
\caption{Branch-number matrix for a cluster of mass 10 million. Values of $N_{ij}$ are given for this 11\textsuperscript{th} ($n=11$) order network, the values of $N_i$ are also given.}
\label{table:branchNum}
\end{table}
\begin{table}
\scriptsize
\setlength{\tabcolsep}{8pt}
\begin{tabular}{c r r r r r r r r r r}
 & \multicolumn{1}{c}{j=2} & \multicolumn{1}{c}{3} & \multicolumn{1}{c}{4} & \multicolumn{1}{c}{5} & \multicolumn{1}{c}{6} & \multicolumn{1}{c}{7} & \multicolumn{1}{c}{8} & \multicolumn{1}{c}{9} & \multicolumn{1}{c}{10} & \multicolumn{1}{c}{11}\\
i=1 & 1.29 & 3.24 & 8.58 & 22.30 & 63.17 & 170.85 & 407.87 & 1,694.27 & 2480.33 & 7287 \\
2 & 0 & 1.09 & 2.87 & 7.59 & 21.65 & 58.29 & 141.43 & 591.87 & 887.33 & 2554 \\
3 & 0 & 0 & 1.04 & 2.74 & 7.97 & 21.49 & 52.10 & 219.73 & 310.67 & 940 \\
4 & 0 & 0 & 0 & 0.99 & 2.95 & 8.32 & 20.43 & 81.53 & 120.67 & 366 \\
5 & 0 & 0 & 0 & 0 & 1.06 & 3.04 & 7.02 & 33.47 & 52.33 & 138 \\
6 & 0 & 0 & 0 & 0 & 0 & 1.15 & 2.82 & 11.73 & 15.00 & 44 \\
7 & 0 & 0 & 0 & 0 & 0 & 0 & 0.98 & 4.60 & 5.67 & 23 \\
8 & 0 & 0 & 0 & 0 & 0 & 0 & 0 & 1.27 & 2.33 & 5 \\
9 & 0 & 0 & 0 & 0 & 0 & 0 & 0 & 0 & 1.67 & 4 \\
10 & 0 & 0 & 0 & 0 & 0 & 0 & 0 & 0 & 0 & 1 \\
\end{tabular}
\caption{Branching-ratio matrix for a cluster of mass 10 million. Values of $T_{ij}$ are for this 11\textsuperscript{th} ($n=11$) order network.}
\label{table:branchRatio}
\end{table}

As noted previously it has been shown that river drainage networks are well approximated by Tokunaga branching statistics. We now compare our branching statistics for the percolation cluster with branching statistics for two river networks. Peckham \cite{Peckham1995} has determined branching statistics for the Kentucky River basin in Kentucky and the Powder River basin in Wyoming. Both of these river networks are 8th order $(n=8)$. For the Kentucky River basin, the bifurcation ratio is $R_b=4.6$, the length-order ratio is $R_r=2.5$, the fractal dimension is $D=1.67$ and the Tokunaga branching constant is $c=2.5$. For the Powder River basin the bifurcation ratio is $R_b=4.7$, the length-order ratio is $R_r=2.4$, the fractal dimension is $D=1.77$, and the Tokunaga branching constant is $c=2.5$. For comparison, the values we obtained for our percolation cluster were $R_b=4.581$, $R_r=2.658$, $D=1.557$, and $c=2.642$. These values are summarized in Table \ref{table:comparison}. Tokunaga self-similar branching networks are universally accepted as the extraction mechanism for water in drainage networks. It has been argued that these networks provide optimum removal of water in landscapes \cite{Rodriguez-Iturbe1997}. We argue that Tokunaga scaling of our invasion percolation network provides an explanation for the efficient extraction of tight shale oil and gas in a high volume fracking injection. 

The concept of Tokunaga self-similar side branching was independently introduced into the physics literature by Vannimenus and Viennot \cite{Vannimenus1989}. Using their approach Ossadnik \cite{Ossadnik1992} studied the branching statistics of diffusion-limited aggregation (DLA) clusters. We note that Ossadnik used a slightly different branch labeling scheme. Average statistics were obtained for 47 DLA clusters each with $10^6$ particles. On average the clusters were 11th order $(n=11)$ fractal trees. The average bifurcation ratio was $R_b=5.15$ and the average length-order ratio was $R_r=2.86$. From Eq. \eqref{eq:TfractalDim} the corresponding fractal dimension is $D=1.56$. An excellent correlation with Tokunaga network statistics was also found using Eq. \eqref{eq:tokunaga} with $c=2.7$. These values are compared with our values given above in Table \ref{table:comparison}. Clearly the branching statistics for our invasion percolation cluster are quite similar to the branching statistics of DLA clusters. It is interesting to note that alternative invasion percolation \cite{Stark1991} and DLA \cite{Masek1993} models have been published to explain the Tokunaga statistics of drainage networks.
\begin{table}
\setlength{\tabcolsep}{12pt}
\begin{tabular}{l c c c c c}
 & n & $R_b$ & $R_r$ & $D$ & $c$ \\
Invasion Percolation (This Paper) & 11 & 4.581 & 2.658 & 1.557 & 2.642 \\
Kentucky River Basin \cite{Peckham1995} & 8 & 4.6 & 2.5 & 1.67 & 2.5 \\
Powder River Basin \cite{Peckham1995} & 8 & 4.7 & 2.4 & 1.77 & 2.5 \\
DLA \cite{Ossadnik1992} & 11 & 5.15 & 2.86 & 1.56 & 2.70
\end{tabular}
\caption{Comparison of branching statistics for our invasion percolation cluster (this paper), with branching statistics for two river networks \cite{Peckham1995}, and with branching statistics for DLA clusters \cite{Ossadnik1992}. The network order $n$, branching ration $R_b$, length-order ratio $R_r$, fractal dimension $D$, and Tokunaga branching constant $c$ defined in Eq. \eqref{eq:tokunaga} are given.}
\label{table:comparison}
\end{table}

\section{Discussion}
It should be emphasized that our planar 2D invasion percolation model does not include many of the known features of actual reservoirs (three dimensions, anisotropic nonhomogeneous stress fields, natural fractures, joint sets, faults, etc.); however, it does provide a simple framework in which these features can incorporated. We will give just a few of the ways these features can be incorporated. The procedure illustrated in Fig. \ref{fig:procedure} can easily be extended to a cubic lattice to allow for simulation of three dimensional. Microseismic activity associated with high volume fracking tends to be confined to a horizontal layer which is attributed to the vertical stress being the maximum compressive stress. This anisotropic stress can be built into the random numbers by choosing random numbers in the range 0 to 1 for the vertical direction and random numbers in a smaller range, say 0 to 0.25 for the horizontal direction so that fractures are 4 times more likely to propagate in the horizontal direction than the vertical direction. Additionally, if the local stresses within the reservoir are known, the random numbers could be modulated based upon the local stress. Natural fractures, joint sets and faults can be included by introducing regions with lower random numbers. In the case of a faults, all the bonds lying in a plane could have their strength values reduced by a factor of 10 so that once the growing fracture network reaches the fault, fracture growth is likely to be confined to the fault plane. These and other additions will be the subject of future work.

One of the basic assumptions in most percolation papers is that the bond strengths are assigned independently so that the strengths are uncorrelated. However, the granular permeability of sandstones has been shown to have long-range correlations \cite{Turcotte1997}. Prakash et al. \cite{Prakash1992} solved an invasion percolation problem with spatially correlated occupancy variables. The correlations were obtained using fractional Brownian motion. These studies showed that scaling exponents depend on the Hurst exponent of the assumed correlation. Herrmann and Sahimi \cite{Herrmann1993} studied invasion percolation with radially dependent occupancy variables.

It is clearly desirable in future studies of invasion percolation associated with fracking to include spatially correlated bond strengths. However the applicable correlation laws applicable to super fracking are far from clear. The permeability in tight shales is associated with natural fractures generated by oil and gas generation. Studies of the statistics of natural fractures \cite{Engelder2009} indicate that they are quasi periodic with spacings of 0.1 to 1.0 meters. The deposition of carbonates in the natural fractures generates the tight shales subjected to super fracking. The spatial correlation of the resistance to opening of the sealed fractures by high pressure fluid injection is not established.

Before considering the many possible additions to this model, we felt it valuable to first characterize the simplest version of the model. This allowed us to determine that this model is indeed unique from other percolation models and determine some of the qualitative aspects of this model that are relevant for fracking.

We have shown in Figure \ref{fig:burstDist} the power-law scaling of the bursts of cluster growth in our model. We believe this provides an explanation for the power-law scaling of the microseismic fractures observed in high volume fracking. However, we would expect different power-law exponents due to the geometrical limitations of our 2D model. The network of propagating fractures as indicated by microseismicity \cite{Maxwell2011} certainly resembles our expanding percolating cluster.

We have also quantified the branching statistics of our evolving percolation cluster. Specifically, we obtain excellent Tokunaga self similar scaling as shown in Figure \ref{fig:tokunaga}. This scaling is quantitatively similar to the scaling of river networks. These networks are recognized as an optimal geometry for extracting the water from rainfall in a river basin. The landscape associated with a drainage topography is similar to the distribution of fracture permeabilities associated with a tight-shale reservoir. We argue that the Tokunaga scaling of our invasion percolation network provides an explanation for the efficient extraction of oil and gas by the fracture network generated in a high volume fracking injection.

\section{Acknowledgements}
The authors would like to acknowledge valuable discussions with William Klein and Armin Bunde. We would also like to acknowledge the assistance of Mark Yoder and Eric Heien during code development. Additionally, we would like to thank the two anonomous reviewers for their thorough critique of our manuscript. The research of JQN and JBR has been supported by a grant from the US Department of Energy to the University of California, Davis \#DE-FG02-04ER15568.

\clearpage
\bibliography{IPPaper}

\end{document}